\documentclass[sigconf]{acmart}

\AtBeginDocument{%
  \providecommand\BibTeX{{%
    \normalfont B\kern-0.5em{\scshape i\kern-0.25em b}\kern-0.8em\TeX}}}

\copyrightyear{2025}
\acmYear{2025}
\setcopyright{acmlicensed}\acmConference[CHI '25]{CHI Conference on Human Factors in Computing Systems}{April 26-May 1, 2025}{Yokohama, Japan}
\acmBooktitle{CHI Conference on Human Factors in Computing Systems (CHI '25), April 26-May 1, 2025, Yokohama, Japan}
\acmDOI{10.1145/3706598.3713357}
\acmISBN{979-8-4007-1394-1/25/04}

\newcommand{\sys}{{Codellaborator}{}}

\newcommand{\revise}[2][black]{\textcolor{#1}{#2}}

\definecolor{lightgrey}{rgb}{0.8,0.8,0.8}
\definecolor{white}{rgb}{1.0,1.0,1.0}

\usepackage{makecell}
\usepackage{longtable}
\usepackage{listings}
\usepackage{xcolor}
\usepackage{hyperref}
\usepackage{hyperxmp}

\usepackage{array}
\usepackage{amsmath}
\usepackage{microtype}
\usepackage{graphicx}
\usepackage{float}
\usepackage{wrapfig}
\usepackage{xspace}
\usepackage{enumitem}
\setlistdepth{9}

\usepackage{lipsum}
\usepackage{todonotes}
\usepackage{tikz}

\begin{document}

\title[Exploring and Evaluating Proactive AI Programming]{Assistance or Disruption? Exploring and Evaluating the Design and Trade-offs of Proactive AI Programming Support}

\author{Kevin Pu}
\email{jpu@dgp.toronto.edu}
\affiliation{
    \institution{University of Toronto}
    \country{Canada}
}

\author{Daniel Lazaro}
\email{d.lazaro@mail.utoronto.ca}
\affiliation{
    \institution{University of Toronto}
    \country{Canada}
}

\author{Ian Arawjo}
\email{ian.arawjo@umontreal.ca}
\affiliation{
    \institution{Université de Montréal}
    \country{Canada}
}

\author{Haijun Xia}
\email{haijunxia@ucsd.edu}
\affiliation{
    \institution{University of California San Diego}
    \country{USA}
}

\author{Ziang Xiao}
\email{ziang.xiao@jhu.edu}
\affiliation{
    \institution{Johns Hopkins University}
    \country{USA}
}

\author{Tovi Grossman}
\email{tovi@dgp.toronto.edu}
\affiliation{
    \institution{University of Toronto}
    \country{Canada}
}

\author{Yan Chen}
\email{ych@vt.edu}
\affiliation{
    \institution{Virginia Tech}
    \country{USA}
}

\renewcommand{\shortauthors}{Pu et al.}
\begin{abstract}
    AI programming tools enable powerful code generation, and recent prototypes attempt to reduce user effort with proactive AI agents, but their impact on programming workflows remains unexplored.
    We introduce and evaluate Codellaborator, a design probe LLM agent that initiates programming assistance based on editor activities and task context. 
    We explored three interface variants to assess trade-offs between increasingly salient AI support: prompt-only, proactive agent, and proactive agent with presence and context (Codellaborator). 
    In a within-subject study ($N=18$), we find that proactive agents increase efficiency compared to prompt-only paradigm, but also incur workflow disruptions. 
    However, presence indicators and interaction context support alleviated disruptions and improved users' awareness of AI processes. 
    We underscore trade-offs of Codellaborator on user control, ownership, and code understanding, emphasizing the need to adapt proactivity to programming processes. 
    Our research contributes to the design exploration and evaluation of proactive AI systems, presenting design implications on AI-integrated programming workflow.
\end{abstract}

\ccsdesc[500]{Human-centered computing~Interactive systems and tools}
\ccsdesc[500]{Human-centered computing~Empirical studies in HCI}

\keywords{Proactive AI; Programming Assistance; Human-AI Interaction}

\maketitle

\section{Introduction}
Large language models (LLMs) have enabled generative programming assistance tools to provide powerful in-situ developer support for novices and experts alike \cite{vaithilingam2022expectation, murillo2023exploreaccelerate, majeed2024novice, majeed2023intro, Khojah2024BeyondCG}.
Many existing LLM-based programming tools rely on user-initiated interactions, requiring prompts or partial code snippets as input to provide sufficient context and to trigger help-seeking \cite{vaithilingam2022expectation, Nam2023InIDEGI, Mcnutt2023OnTD, ross2023programmerassistant, jiang2022genline}.
These systems offer help in the form of code output and natural language explanations to assist users with coding tasks. 
However, research indicates that users invest considerable effort in formulating prompts, interpreting responses, assessing suggestions, and integrating results into their code \cite{Mozannar2022ReadingBT, vaithilingam2022expectation, jiang2022genline}. 

To alleviate the user intent specification costs, recent AI programming tools and designs aim to become more autonomous, allowing the system to initiate interaction and provide proactive assistance.
Tools like Github Copilot \cite{githubcopilot} and Visual Studio’s IntelliCode \cite{Svyatkovskiy2020intellicode} offer the auto-completion feature, which fills the code line as the user is typing, to alleviate prompt engineering efforts and proactively provide in-situ support.
However, the AI tool's generated code is not always accurate and the output still requires significant user effort to verify \cite{vaithilingam2022expectation, Vaithilingam2023intellicode, Mozannar2022ReadingBT, liang2024usabilityCopilot}.
To address this, commercial and research prototypes constructed intelligent AI agents with distinct focuses on the \revise{\emph{timing of assistance}}, \revise{\emph{the representation of the agent}}, and \revise{\emph{the scope of the interaction context}}, aiming to work with the user as collaborators and tackle coding tasks autonomously and preemptively.
These approaches involve advancing on the timing of auto-completion feature to proactively make intelligent file changes with an AI caret in the code editor \cite{cursorcopilot++}, representing the AI's presence in the editor with an automated AI cursor \cite{geniusbydiagram}, \revise{or grounding interactions context in different scopes, such as conversational dialogue \cite{copilotX, ross2023programmerassistant, ReplitAI}, specific code lines \cite{jiang2022genline, githubcopilot}, or an agent-managed workspace to tackle software engineering tasks autonomously \cite{devinAISWE, yang2024sweagentagentcomputerinterfacesenable, wang2024opendevinopenplatformai}}.
However, the effects of these new designs of system-driven programming assistance on the human workflows, compared to the existing user-initiated paradigm, remain to be explored.
Visions of more ``\textit{proactive}\footnote{\revise{In this paper, we refer to system-initiated assistance in the programming environment as ``proactive'' programming support.}}'' AI programmers additionally raise questions about the potential for harm. 
Researchers have raised on concerns that excessive automation without proper human control can lead to unreliable and unsafe systems \cite{Shneiderman_2020_hcai}, and thus some AI systems deliberately avoid proactive AI assistance \cite{ross2023programmerassistant}.
Therefore, there remain the questions of \revise{\emph{when}} should AI programming tools provide proactive support, \revise{\emph{how} should the support be delivered, and \emph{where} should the user interact with such support.} \revise{Subsequently}, what are the effects of a proactive AI programming tool on user experience? \revise{In which programming processes and tasks} is proactivity helpful, and where might it be harmful? 

\revise{This research explores the design space of proactive AI programming tools in three dimensions -- the timing of assistance, the representation of the AI programming tool, and the scope of interaction context.} We then evaluate the effects of this human-AI interaction paradigm on software engineering practice, illustrating the advantages and drawbacks to provide insights for future designs. We were guided by these research questions:
\begin{itemize}
    \item \textbf{RQ1:} How can we design \revise{proactive assistance in an AI programming tool} to reduce user effort?
    \item \textbf{RQ2:} What are the benefits and drawbacks of a proactive AI programming tool compared to user-initiated systems?
    \item \textbf{RQ3:} In which programming processes and task contexts can proactivity be helpful, and where can it be harmful?
\end{itemize}

\begin{figure*}[h]
    \centering
    \includegraphics[width=0.76\textwidth,keepaspectratio]{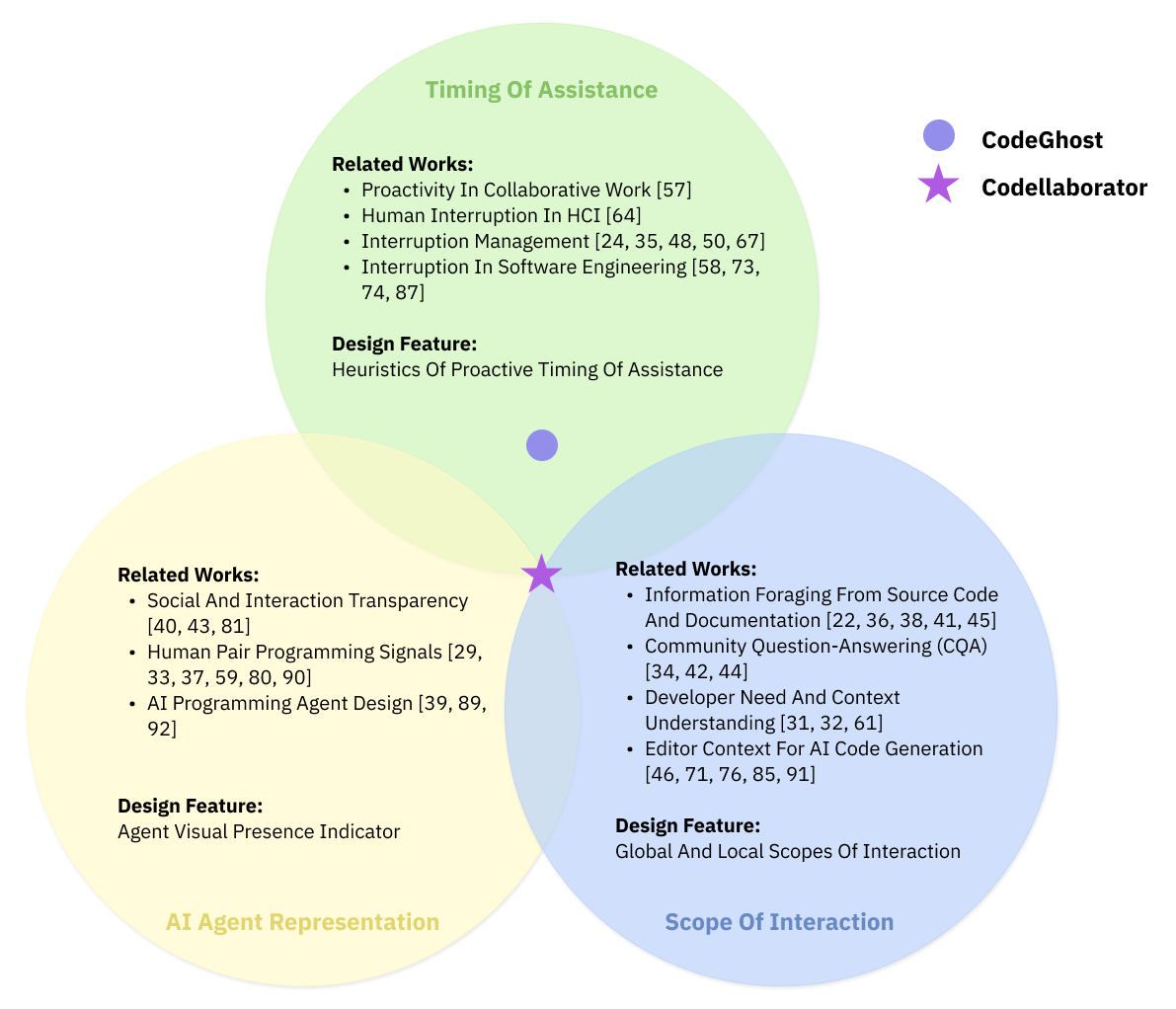}
    \caption{\revise{\textbf{System Design Dimensions.} \textnormal{\sys{} explores three design dimensions. Each dimension is motivated by relevant theories under the framework of effective human collaboration; 
    The \emph{Timing of Assistance} (DG1) design takes inspiration from works in proactivity \cite{kim2020psychological} and interruption management in human collaboration \cite{mcfarlane2002scope, bailey2000measuring,horvitz2001notification,czerwinski2000instant, Miyata1986, iqbal2005towards} and specifically the software engineering context \cite{Solingen1998InterruptsJA, Ko2007InformationNI, parnin2010cues-resuming, parnin2011resumption-strategies}.  
    The \emph{AI Agent Representation} (DG2) dimension draws from literature in social and interaction transparency \cite{stuart2012social, erickson2003social, gutwin2004group}, pair programming signals \cite{cockburn2000costs, williams2003pair, Begel_pair, lee2017exploring, stein2004another, dourish1992awarness_workspace}, and existing AI programming agent designs \cite{wang2024opendevinopenplatformai, yang2024sweagentagentcomputerinterfacesenable, ehsan2021expanding}.
    The \emph{Scope of Interaction} (DG3) design is informed by works in information foraging in coding \cite{Fleming2013AnIF, Deline2006CodeTU, DualaEkoko2010TheIG,Aghajani2019SoftwareDI, Horvath2021UnderstandingHP}, community question-answering tools \cite{Goldman2008CodetrailCS, cordeiro2012contextQA, Hartmann2011HyperSourceBT}, research in understanding developer need and context \cite{chen2016towards, mamykina2011design, chen2017codeon}, and editor context scopes for AI code generation \cite{yan2024ivie, Vaithilingam2023intellicode, Nam2023InIDEGI, ross2023programmerassistant, horvath2024meta-manager}.
    \sys{} is a design probe that explores designs that intersect each dimensions and evaluates their impacts on users' programming experiences. Two other probes are described in Section \ref{Evaluation:study-design} (DG4). CodeGhost explores the effect of proactive timing heuristics without the agent representation and scope of interaction designs, and PromptOnly is the baseline condition that does not inherit any of the design explorations and thus is not illustrated on the diagram. Please note that we categorized each reference based on the primary aspect it informs within our framework, but some of the cited works may span multiple dimensions and we acknowledge their relevance across overlaps.}}}
    \label{fig:theory}
\end{figure*}

To answer these research questions, 
we \revise{incorporate} theories of \revise{interruption management, social transparency, and help-seeking behavior in programming (Fig.\ref{fig:theory}) to identify specific design rationales for each dimension}.
Informed by prior literature, we develop \sys{}, a technology probe \cite{hutchinson2003techprobe} that employs an AI programming agent providing \emph{proactive timings of assistance} to explore different forms of human-AI programming collaboration.
\sys{}'s proactive abilities allow it to initiate interaction via messages (Fig.\ref{fig:ui}.a,e) in response to various user activities in the coding environment, and also to commit code edits directly in the editor (Fig.\ref{fig:ui}.c). 
To mitigate potential disruptions, we derive three design rationales for the timing to introduce assistance and operationalize them into six design principles in the context of a coding task and an editor environment (Table \ref{table:proactive-features}).
To \revise{evaluate designs of \emph{AI agent representations} in the editor}, we additionally implemented \textit{presence} features for the AI agent. 
In \sys{}, the agent presence is represented by a cursor and caret (Fig.\ref{fig:ui}.d,b), capable of autonomous movement around the editor, signaling its action, status, and attention focus. 
To \revise{evaluate different \emph{scopes of interaction}, both the agent and the user can utilize global chat messages \revise{on the side-panel}, or initialize locally-scoped threads of conversations called ``breakouts'', anchored to\revise{specific locations in the editor as context} (Fig.\ref{fig:ui}.a)}. 

To study the impact of proactive support in an autonomous coding agent on human-AI collaborative programming workflows, we conducted a within-subject experiment using three versions of \sys{} with 18 participants.
In the PromptOnly condition, the ablated system only responds to user prompts and in-line code comments, similar to ChatGPT \cite{chatgpt} and Github Copilot \cite{githubcopilot} with low to no proactive features. 
In the CodeGhost condition, the system proactively initiates interactions and assistance, but \revise{the interactions are constrained in the global context of chat messages and direct code changes, with no visual representation of the agent.} 
In the \sys{} condition, all of the agent's visual representation, \revise{scopes of interaction}, and proactive timing features are utilized. 

Our study showed that, through the heuristic-based timing to provide contextualized assistance at task boundaries, the CodeGhost condition reduced the time users took to comprehend system responses compared to the PromptOnly condition.
But it also caused workflow disruptions and diminished users' awareness of AI's actions, as participants reported a lack of clear signals for agent interaction and working context. 
In contrast, the \sys{} condition, with its agent visual presence and \revise{flexible context scope}, significantly lessened these disruptions and improved users' awareness of the AI, leading to a user experience more akin to collaborating with a partner than using a tool.
\revise{Participants felt ambivalent to adopt highly proactive programming assistants. Many embraced the efficiency and capability to allow developers to focus on high-level designs rather than low-level work, but some participants experienced a loss of code understanding, expressing concerns on maintainability and extendability of the code artifact.}

In the discussion, we summarize our findings and propose five design implications for proactive assistance in human-AI programming.
Through these findings, we present a deeper understanding of the impacts of proactive AI support on programming experience and identify key areas that require further research.
In this work, we contribute:
\begin{itemize}
    \item A design exploration to enable different interaction timings, \revise{visual representations, and interaction scopes} of proactive assistance that expand upon existing AI programming systems.
    \item \sys{}, as a technology probe that implements a proactive AI agent to study in-situ assistance and communication in programming support.
    \item A empirical study to assess the impact of proactive agent support in a code editor, providing design implications for future AI programming tools.
\end{itemize}

\section{Related Work}

\subsection{AI Programming Tools}
\label{RW:AI_Programming_Tools}
\revise{AI-assisted programming tools are increasingly integrated in developers' workflows.}
However, even tools that prioritize productivity, such as Github Copilot \cite{githubcopilot}, do not consistently demonstrate a significant improvement over traditional code completion tools such as IntelliSense \cite{intellisense, vaithilingam2022expectation, Khojah2024BeyondCG}. 
\revise{Existing research to improve AI programming support has focused on improving the quality and usability of the code generation.}
For example, recent works made advancements in structuring code generation \cite{Yen2023CoLadderSP}, expanding support to specific task domains (e.g., data analysis) \cite{Mcnutt2023OnTD}, or highlighting high-probability tokens to reduce uncertainties \cite{Vasconcelos2023GenerationPA}.
However, AI-generated assistance could result in discrepancies with the user's expectations, creating barriers to interpreting and utilizing the code output, or even steering the AI in the desired direction in the first place \cite{vaithilingam2022expectation}.
Another recurrent concern is the potential mismatch in expertise levels between developers and AI agents, leading to reduced productivity in pair programming scenarios~\cite{Ma2023IsAT}.
\revise{To improve the usability of generated code, researchers enhanced the context within the \emph{scope of interaction} in the code editor.
For example, Yan et al. proposed Ivie, which generates visible explanations positioned adjacent to the code \cite{yan2024ivie}. Similarly, recent systems improved the discoverability of the code suggestion scope \cite{Vaithilingam2023intellicode} or provided more in-IDE code contexts to scaffold user understanding \cite{Nam2023InIDEGI}.
Alternatively, some tools integrates dialogue-based interactions to enable holistic queries at the editor level and enhanced interaction histories \cite{copilotX, ross2023programmerassistant, ReplitAI}.
\revise{Expanding the scope further, Meta-Manager enhances developers' sensemaking by collecting and organizing meta-information, such as code provenance and design rationale, making it easier to answer complex questions about the code base \cite{horvath2024meta-manager}.}
However, it remains unclear whether these approaches to scoping human-AI collaboration will be effective in a system-initiated paradigm and further research is needed to investigate the impact of different scopes of system-initiated actions on developer experience.
}

\revise{Existing AI programming tools predominantly operate within the command-response paradigm, where the user triggers help-seeking and obtains generated code and explanations.}
An emerging approach to enhance AI programming is leveraging LLMs' generative power to build intelligent programming agents that proactively support users and autonomously complete tasks.
Efforts to develop proactive tools that provide automatic support have been explored across various domains, including personalized notifications for weather or calendars \cite{sun2016contextualintent, sarikaya2017digitalassistant}, health and fitness interventions \cite{schmidt2015fitness, rabbi2015healthfeedback}, and support for office workflows \cite{Baym2019clippy, Jacobs2019BeyondCC, matejka2011ambient}. 
Well-designed, effective invocation of systems' proactive assistance can lower the cost of user manipulation, resolve uncertainties preemptively, and lead to unintentional learning of the system's functionalities \cite{matejka2011ambient, horwitz1999mixedinitiative, Sawyer_2014_learning}.
Meanwhile, poorly designed proactive assistance can lead to negative user experiences, diminished control \cite{MORADIDAKHEL2023copilotliability, meurisch2020proactiveexpectation, barkhuus2003losecontrol}, and in some cases, rendering the tool ineffective \cite{meurisch2020proactiveexpectation, Baym2019clippy, Jacobs2019BeyondCC}.
While general guidelines on designing mixed-initiative interfaces and human-AI interactions have been established \cite{horwitz1999mixedinitiative, amershi2019haiprinciple}, it is uncertain how the design principles can translate to concrete system designs under the context of LLM-assisted programming.
For instance, while the \emph{timing of assistance} is one key metric of human-AI interaction design, existing AI programming systems almost always provide immediate response upon output generation without considering interruption to the user's workflow.

Recent research and commercial prototypes have explored many \revise{\emph{representation of the AI agent} to facilitate proactive support in programming.}
Some approaches include expanding on the auto-completion feature to proactively make intelligent file changes with an AI caret in the code editor \cite{cursorcopilot++}, or manifesting the AI's presence in the editor with an automated AI cursor \cite{geniusbydiagram} that mimic the user's workflow and automate repetitive tasks.
Some tools take an additional step towards fully autonomous AI and construct a group of AI agents capable of composing the task plan with executable steps and hosting its own workspace with code editor, console, and web browser to autonomously tackle software engineering tasks in response to a single user prompt \cite{devinAISWE, wang2024opendevinopenplatformai, yang2024sweagentagentcomputerinterfacesenable}.
However, the accuracy of the task completion suggests that the fully autonomous agent might not yet be fully scaled to real-life programming tasks. For example, based on evaluation on SWE-bench \cite{jimenez2024swebenchlanguagemodelsresolve}, a dataset designed to assess AI agent's capabilities in real-world Github issues, SWE-agent and OpenDevin reported 12.5\% and 26.0\% task completion rate.
\revise{This prompts a more balanced design where both the human and the AI agent are engaged in the programming process.}

Further, the impact of proactive programming assistance for human users, as opposed to the current prompt-initiated paradigm, remains to be formally assessed.
Similarly, the resulting benefits and drawbacks to user experience from employing these specific design approaches need to be measured.
Our research not only aims to explore how to effectively design and integrate proactive AI assistance into developers' workflows but also seeks to gain a deeper understanding of the impact on the programming experience through a comprehensive study.

\subsection{\revise{Proactive Assistance and Interruption}}
Designing a proactive AI assistant that enables positive experiences and outcomes is a challenging endeavor. The nuances of effective human-to-human collaboration are still not fully understood and vary greatly depending on the context, making it difficult to craft effective human-AI collaboration paradigms. 
Past research has shown that two factors, i.e., proactivity and interruption, play pivotal roles in shaping the outcomes of team collaborations. 
Prior work in psychology has revealed that proactivity, when effectively managed, can provide positive affective outcomes during collaborative work \cite{kim2020psychological}. However, the current landscape of human-AI collaboration is often characterized by either human-dominant or AI-dominant dynamics. 
In such situations, both human and AI agents operate reactively. \revise{This paradigm often leave the cognitive burdens for human developers due to the expression, sensemaking, and verification process for the code assistance \cite{vaithilingam2022expectation, liang2024usabilityCopilot}.}

Interruption, or \textit{``an event that breaks the coherence of an ongoing task and blocks its further flow, though allowing the primary task to resume once the interruption is removed''} has been a subject of study for decades \cite{mcfarlane2002scope}. 
Numerous studies have highlighted the detrimental effects interruptions can have on users' memory, emotional well-being, and ongoing task execution~\cite{bailey2000measuring,horvitz2001notification,czerwinski2000instant}. 
\revise{Specifically, in the context of software engineering, Solingen et al. defined interruption as a multi-phase process that occurs when a developer stops their planned activities, handles the interruption, then finally recovers by returning to the point in their work at which they were interrupted \cite{Solingen1998InterruptsJA}.
In existing practices, Ko et al. observed developers daily activities and identified multiple interruptions per day, mainly due to communication requests and notifications \cite{Ko2007InformationNI}.
Parnin et al. also found that developers spend significant time rebuilding context after each interruption, creating ``resumption lag'' which increases errors and frustration \cite{parnin2010cues-resuming, parnin2011resumption-strategies}.}
To mitigate the challenges posed by interruptions, we drew insights from psychology and behavioral science to foster collaborations that would be perceived as less disruptive.
For instance, as the perceived level of disruption is influenced by a user's mental load at the time of the interruption~\cite{czerwinski2000instant,bailey2000measuring,bailey2001effects}, we designed interactions that were aware of a user's working context before generating notifications.
Furthermore, prior works have highlighted that people experience varying degrees of disruption during different sub-tasks~\cite{horvitz2001notification,czerwinski2000instant, Miyata1986, iqbal2005towards}. 
We apply these principles when designing the timing of service of our probe to adopt a proactive collaborator role at moments when the programming task context was most appropriate.

\subsection{Help-Seeking and Collaboration in Programming}
Our research was additionally informed by existing research on collaboration during software engineering, specifically help-seeking behaviors and pair programming.

\subsubsection{Help-Seeking in Programming}
\revise{Developers often forage information from the code itself to resolve their issues, during processes like debugging \cite{Fleming2013AnIF}. To facilitate this process, researchers have built tools to scaffold navigation and understanding the source code \cite{Deline2006CodeTU}.
Documentation is another source developers rely on to find assistance.
However, studies have identified challenges for users to pinpoint relevant information and to maintain the documentation in an up-to-date state \cite{DualaEkoko2010TheIG,Aghajani2019SoftwareDI}.
Annotations on documentation, as demonstrated in Adamite, can support comprehension and foster collaboration by addressing gaps in traditional documentation \cite{Horvath2021UnderstandingHP}.}
To seek more targeted help, Community Question-Answering (CQA) websites, such as Stack Overflow \cite{stackoverflow}, also allow developers to post questions but also archive answers for future reference, a concept rooted in Answer Garden's creation of an ``organizational memory''~\cite{ackerman1996answer,ackerman1998augmenting}.
\revise{To more seamlessly connect developer's working context to help seeking, researchers have connected the integrated development environment (IDE) with web browser \cite{Goldman2008CodetrailCS}, web-based Q\&A search \cite{cordeiro2012contextQA}, and annotated the source code with browser histories \cite{Hartmann2011HyperSourceBT}.}
However, many questions that are well-suited for an intelligent agent are misaligned with the design of CQA websites. 
A previous study utilized a ``hypothetical intelligent agent'' as a probe to understand developers' ideal help-seeking needs~\cite{chen2016towards}. 
The findings, along with other studies' results, highlighted several limitations of CQA sites, including delayed feedback, lack of context, and the necessity for self-contained questions~\cite{mamykina2011design}.
Consistent with this, prior work has advocated for systems that intuitively captured a developer's context and used it to enable developers to \revise{identify the scope of assistance by selecting} a code snippet, asking the system to ``please refactor this'', and promptly receiving pertinent responses~\cite{chen2017codeon}.
\revise{Like described in Section \ref{RW:AI_Programming_Tools}, existing AI programming tools often employ different scopes of interaction with the intelligent assistant, translating the consideration of interaction context scope from human help-seeking to human-AI collaboration.}
Inspired by this research, we designed our probe to \revise{allow users to seek help with different granularity of support, receiving assistance on the overall codebase via a global chat interface and on specific code snippets via localized conversation threads. This way, we can evaluate the impacts of different scope of interactions in a more system-initiated programming paradigm.}

\subsubsection{Pair Programming}
Pair programming is a paradigm where two users collaborate in real-time while at a single computer, with one user writing the code (i.e., the driver) and the other reviewing the code (i.e., the observer) \cite{cockburn2000costs}. 
Pair programming has been shown to lead to better design, more concise code, and fewer errors within approximately the same person-hours~\cite{williams2003pair, Begel_pair}.
Other research has reported that these benefits may have been due to the awareness of another's focus within the code, which can be invaluable for problem-solving~\cite{lee2017exploring}.
For example, Stein and Brennan found that when novices observed the gaze patterns of expert programmers during code reviews, they pinpointed bugs faster~\cite{stein2004another}.
However, the most prominent challenges associated with pair programming include cost inefficiency, scheduling conflicts, and personality clashes~\cite{Begel_pair}.
Facilitating visible presence and actions between collaboration partners has previously demonstrated its efficacy in physical workspaces \cite{dourish1992awarness_workspace}.
Our design probe loosely adopted the pair programming paradigm where the AI agent and the user can adapt and exchange the roles of the driver and the observer.
We also implemented visible presence and clear context information to enhance mutual awareness between the user and the AI.

\section{Design Goals}
To investigate the effects of a proactive AI programming agent on user workflow, we used a technology probe ---  an \textit{instrument that is deployed to find out about the unknown—returning with useful or interesting data} \cite{hutchinson2003techprobe} --- to explore the design space of the \revise{timing, representation, and scope of interaction} (RQ1) with the following design considerations:

\begin{itemize}
    \item \textbf{\revise{DG1: Establish heuristics for timely proactive AI assistance}} by anticipating programmer needs based on editor activities and offering suggestions, insights, or corrections to support their tasks.
    \item \textbf{\revise{DG2: Represent AI agent's presence}} using visible cues to indicate its actions, intentions, and decision-making processes, enhancing the user's awareness of the agent's assistance.
    \item \revise{\textbf{DG3: Provide flexible scopes of interaction} by designing interactions at both a global code editor level and at a local code-line level to meet different abstractions of user need and improve context management.}
    \item \textbf{DG4: Support different mechanisms in the probe and creating different versions of the system} to structure comparisons and evaluations of different designs    
\end{itemize}

\section{The \sys{} Probe}
In this section, we introduce three main components of the \sys{} probe and how they are implemented to achieve our design goals: \revise{timely proactive support, AI agent's visual representation, and multi-level scopes of interaction}, supporting modular comparisons of different mechanisms.
Below we detail the design and implementation of the probe.

\subsection{\revise{Timely Proactive Programming Assistance}}
Timing services based on context is a key consideration in AI system design \cite{amershi2019haiprinciple}.
To explore the design for proactive timings specifically in the context of programming support, 
we adopted a set of findings from research on \textit{interruption management}, and distilled them down to three proactivity design principles \cite{Miyata1986, iqbal2005towards, czerwinski2000instant, bailey2001effects}. 
To operationalize, we instantiated six proactivity features in our design probe to minimize interruptions to the user (DG1), summarized in Table \ref{table:proactive-features}.
The proactive assistance in \sys{} serves to present one design approach that expands upon existing AI programming features, allowing us to investigate the effects of an AI agent equipped with highly proactive capabilities on users' programming workflow.

\textbf{The first principle} states that the most opportune moments for interruption occur during periods of low mental workload~\cite{bailey2001effects,czerwinski2000instant}. 
In our system, we predict low mental workloads when users are not performing actions, such as writing code, moving around the file, and selecting ranges (i.e. when the user is idle).
Since idleness could also mean the user is engaged in thoughts, the AI agent only intervenes after an extended period of inactivity in both editing and cursor movement, which could signal that the user is mentally stuck and needs assistance (Table \ref{table:proactive-features}, 1).
This is a rough estimation to interpret the user's working states, and future works could employ more advanced models to identify the user's cognitive process.

\begin{table*}[]
\renewcommand{\arraystretch}{1.1}
\begin{tabular}{lll}
\hline
\textbf{User Action} &
  \textbf{System Reaction Trigger} &
  \textbf{Possible Action Space} \\ \hline
\multicolumn{3}{l}{\textbf{Design Rationale 1:} intervene at moments of low mental workload \cite{bailey2001effects, czerwinski2000instant}} \\ \hline
\multicolumn{1}{l|}{\begin{tabular}[c]{@{}l@{}}1. User has been idle\\ (no code edit, \\ caret movement, \\ or selection change)\end{tabular}} &
  \multicolumn{1}{l|}{\begin{tabular}[c]{@{}l@{}}Initial idle threshold is 30 seconds.\\ If user ignores and maintains idle,\\ add 30 seconds to the threshold.\end{tabular}} &
  \begin{tabular}[c]{@{}l@{}}1. If the user is on an empty \\ or trivial line (e.g. pass), no response\\ 2. Offer help via message\end{tabular} \\ \hline
\multicolumn{3}{l}{\begin{tabular}[c]{@{}l@{}}\textbf{Design Rationale 2:} intervene at task boundary (i.e. when completed one subtask and formulating the next) \cite{iqbal2005towards, Miyata1986}\end{tabular}} \\ \hline
\multicolumn{1}{l|}{\begin{tabular}[c]{@{}l@{}}2. User has completed \\ a block of code\end{tabular}} &
  \multicolumn{1}{l|}{\begin{tabular}[c]{@{}l@{}}User outdents from a code scope in Python \\ (e.g. an if- statement, loop, or function).\end{tabular}} &
  \begin{tabular}[c]{@{}l@{}}1. If block is insignificant, no response\\ 2. Notify user of code issues\\ 3. Suggest optimization to user\\ 4. Adds documentation in editor\end{tabular} \\ \hline
\multicolumn{1}{l|}{\begin{tabular}[c]{@{}l@{}}3. User has executed \\ the program\end{tabular}} &
  \multicolumn{1}{l|}{\begin{tabular}[c]{@{}l@{}}User executes the program in editor.\\ Output displays in console.\end{tabular}} &
  \begin{tabular}[c]{@{}l@{}}Acknowledge the code execution. \\ If output contains error, offer to help.\end{tabular} \\ \hline
\multicolumn{1}{l|}{\begin{tabular}[c]{@{}l@{}}4. User has made \\ a multi-line \\ code change\end{tabular}} &
  \multicolumn{1}{l|}{\begin{tabular}[c]{@{}l@{}}User pastes code \\ that's more than 1 line.\end{tabular}} &
  \begin{tabular}[c]{@{}l@{}}1. If change is insignificant, no response\\ 2. Add documentation in editor\\ 3. Notify user of code issues\end{tabular} \\ \hline
\multicolumn{3}{l}{\textbf{Design Rationale 3:} intervene when user is potentially communicating through implicit signals \cite{Nam2023InIDEGI,githubcopilot,copilotX}} \\ \hline
\multicolumn{1}{l|}{\begin{tabular}[c]{@{}l@{}}5. User has made \\ a code comment\end{tabular}} &
  \multicolumn{1}{l|}{\begin{tabular}[c]{@{}l@{}}User starts a newline after \\ a single line or multi-line comment.\end{tabular}} &
  \begin{tabular}[c]{@{}l@{}}1. If nothing to address, no response\\ 2. If the comment describes function, \\ generate code suggestion\\ 3. If posing a question, offer help\end{tabular} \\ \hline
\multicolumn{1}{l|}{\begin{tabular}[c]{@{}l@{}}6. User maintains selection \\ on a range of code\end{tabular}} &
  \multicolumn{1}{l|}{\begin{tabular}[c]{@{}l@{}}Initial selection threshold is 15 seconds. \\ If user ignores and maintains selection, \\ add 15 seconds to the threshold.\end{tabular}} &
  \begin{tabular}[c]{@{}l@{}}1. If insignificant selection, no response\\ 2. Explain the selected code\\ 3. Analyze selection to error-check\end{tabular}
\\ \hline
\end{tabular}
\caption{\textbf{Proactivity features in \sys{}}. \textnormal{The table details the design rationales derived from interruption management literature \cite{Miyata1986, iqbal2005towards, czerwinski2000instant, bailey2001effects} and prior tools \cite{copilotX, githubcopilot, Nam2023InIDEGI}. We designed six proactivity features triggered by user activities. \revise{For each feature, the AI agent evaluates the working context (i.e. the changed code, the user's caret location, and local file content) provided via prompts (Appendix \ref{appendix:prompt}) and decides on one action from the defined list of possible actions, employing necessary tools to make editor changes.}}
}
\label{table:proactive-features}
\end{table*}

\textbf{The second principle} posits that people perceive interventions as less disruptive at the beginning of a task or subtask boundaries \cite{czerwinski2000instant, iqbal2005towards, Miyata1986}.
Task boundaries are defined as when one subtask is completed (evaluation) or when the next subtask begins (goal formulation) \cite{Miyata1986}.
To convert this implication to a system feature, we used event listeners to detect users' task beginnings and boundaries in programming, specifically, when the user completed a block of code (i.e., after outdenting in Python), executed the code, or made a multi-line edit (i.e. pasting a block of code) (Table \ref{table:proactive-features}, 2-4).

Finally, we draw inspiration from existing AI programming tools \cite{githubcopilot, copilotX, Nam2023InIDEGI} and propose \textbf{the third principle}: intervene when users are communicating through implicit signals.
Existing systems don't always use direct messages as the means of human-AI communication.
For example, in Github Copilot \cite{githubcopilot}, creating a new line after a comment prompts the tool to generate code based on the comment content. 
Another example is in Nam et al.'s work, where the system uses the user's current selection in the editor as context to provide code generation \cite{Nam2023InIDEGI}.
While these features demonstrate a low level of proactivity individually, we assimilate the existing designs to enhance the proactivity of \sys{}.

\begin{figure*}[t]
    \centering
    \includegraphics[width=\textwidth]{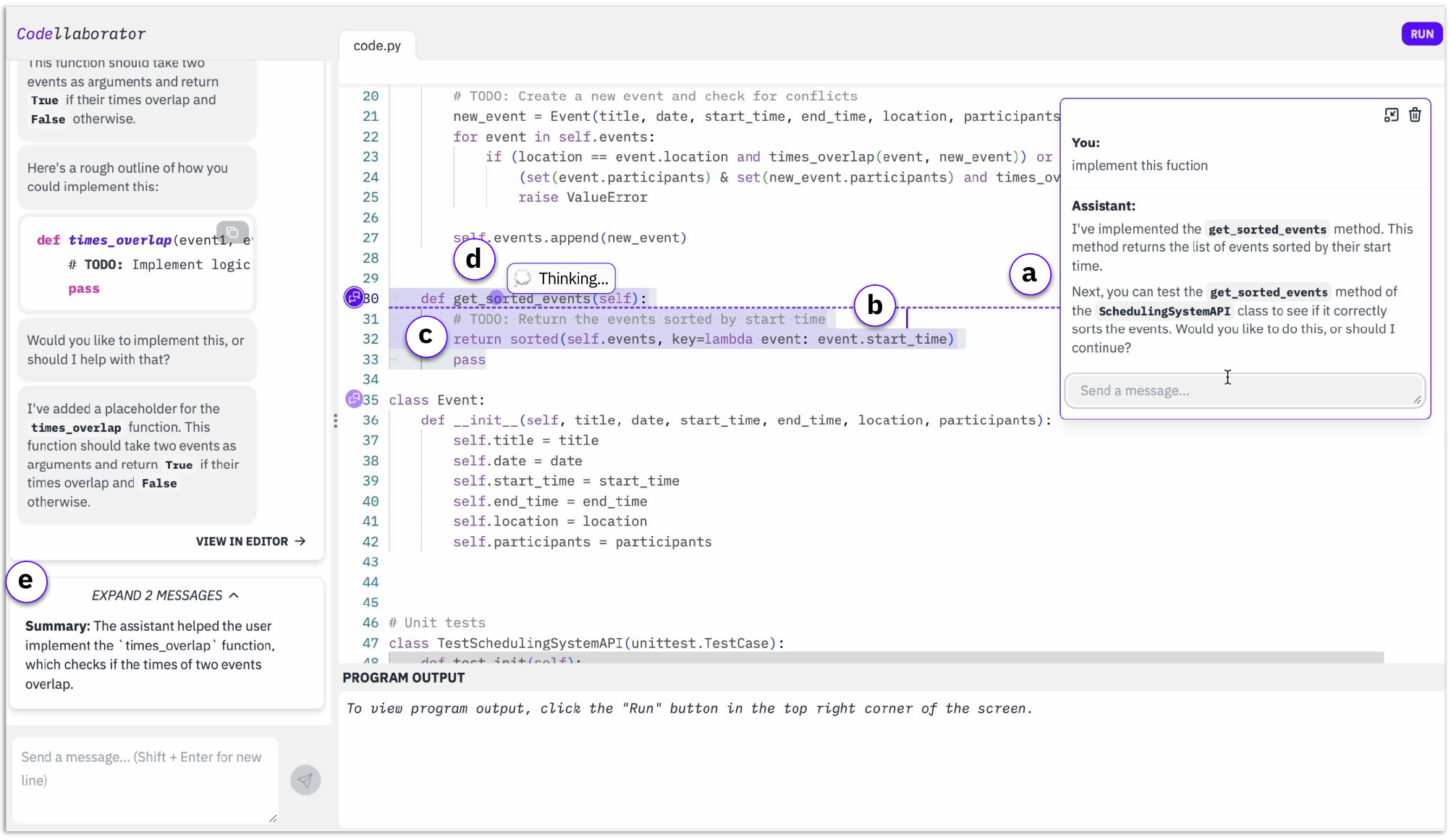}
    \caption{\textbf{\sys{} UI in action.} \textnormal{The user asks the AI agent for help with implementing \texttt{get\_sorted\_events} using a breakout chat \textcircled{\raisebox{-0.5pt}{a}} on line 30. The AI adds code with its caret \textcircled{\raisebox{-0.5pt}{b}} in the editor and replies to the user in the breakout to discuss the next steps. The purple highlight  \textcircled{\raisebox{-0.5pt}{c}} indicates the provenance of the added code and fades away after 5 seconds. The AI agent's cursor displays ``Thinking...'' \textcircled{\raisebox{-0.5pt}{d}} to indicate its working process of generating a response. In the main chat panel to the left, messages are automatically organized by topic with summaries \textcircled{\raisebox{-0.5pt}{e}}. Note that during actual usage, when a breakout is opened, the main chat panel is blurred to alleviate cognitive load and signal context focus switch, but this feature was disabled for UI demonstration purposes.}}
    \label{fig:ui}
\end{figure*}

In each proactivity feature, the AI agent receives the user's caret position and local code context.
It then leverages the LLM to reason, triage, and decide whether to intervene, selecting an option from the defined list of editor actions if deemed necessary.
We also implemented adaptiveness within the AI agent's proactivity. 
For example, each time an idle or selection intervention (Table \ref{table:proactive-features}.1,6) was ignored by the user (i.e. no follow-up interaction with the agent), we imposed a penalty on the action and increased the time threshold to trigger an intervention to make it less frequent in the future to decrease unnecessary interventions.
Additionally, we prioritized the user's initiative and actions over the AI agent's, providing the user with ultimate control. When the user was initiating a conversation with the assistant, we canceled pending AI agent actions and active API requests, so as to not disrupt the user’s train of thought and wait for their updated input. 
\revise{However, to support parallel workflows (e.g., when the user is writing code while the agent proactively modifies other parts of the code), the agent does not cancel its actions if the user is making changes in the editor.}
To enable users to control the amount of visual signals they receive, the chat interface could be fully collapsed, providing more space for the code editor and hiding potentially distracting messages. 

In general, these guidelines lead the probe system to create proactive LLM-based agents for coding support. The agent accounts for the user's current text context and past interactions before evaluating which action to take and when to intervene. When appropriate, the agent can \textbf{proactively} make requests to the LLM and take agent actions to help facilitate direct communication and collaboration with the user.
With this approach, we attempt to improve the intervention timing using both rule-based heuristics and LLM decision-making predictions, constraining the model to take feasible and reasonable actions in a programming assistance context.
We conduct interaction-level analysis in a user study to evaluate the effectiveness of each timing principle and offer design insights for future proactive AI programming systems.

\subsection{\revise{AI Agent's Visual Representation in the Code Editor}}
\revise{To explore the effects of enhanced visual representation, our prototype makes visible the agent's actions, status, and process (DG2).}
\sys{} manifests the AI's presence with visual cues guided by the Social Transparency theory \cite{stuart2012social, erickson2003social} and existing design explorations for programming agents \cite{ehsan2021expanding, cursorcopilot++, geniusbydiagram}.
Specifically, we were inspired by the concept of interaction transparency, which posits that the visibility of the presence of other parties and sources of information in human collaboration can reduce interaction friction \cite{gutwin2004group, erickson2003social}.

To do so, we added a visual caret and cursor in the editor workspace automated by the AI agent. The AI caret indicated its position in the text buffer and moved as the AI agent selected and edited code (Fig.\ref{fig:ui}.b). 
The AI cursor, which moved independently of the caret, serves as an indicator of the AI agent’s “attention” and demonstrates its actions (Fig.\ref{fig:ui}.d). 
For example, to rewrite a block of code, the AI cursor would select a range of code, delete the range, and stream the new code in a typing motion, similar to how a human user would act.
We introduced human elements into the AI agent's actions to elicit the social collaboration heuristics to facilitate better human-AI collaboration \cite{ehsan2021expanding}.

While the AI agent is processing or taking action, a thought-bubble overlay floats adjacent to the cursor and contains an emoji and short text to convey the AI agent's working state (Fig.\ref{fig:ui}.d). 
For example, a writing hand emoji with ``Writing code...'' indicated that the AI agent was writing code in the editor, whereas a tool and laptop emoji with ``Program executing...'' indicated that it was analyzing the program execution output. 
The system also indicated its working progress by streaming a response with a ``pending'' indicator in the chat panel and a ``Loading'' signal on the AI cursor, thus preventing confusion about whether the system was responsive or stalled.
This design enabled users to be aware of system actions even when the chat panel was collapsed, allowing them to decide whether to engage based on the status displayed. 
The option to attend to the textual messages or the visual presence afforded different levels of interaction details, handing users control over the amount of information to be received from the AI agent.

\subsection{\revise{Different Scopes of Interaction}}
\revise{To examine the effects of different scopes of human-AI interaction within the code editor (DG3), \sys{} affords two channels for either the user or the agent to initiate an interaction. 
This was inspired by the literature on help-seeking in programming, where users often need to define the context (i.e. relevant code snippets \cite{chen2017codeon}, code annotations \cite{Horvath2021UnderstandingHP}, or search history \cite{Hartmann2011HyperSourceBT}) to request assistance.
Existing AI programming tools often adopt conversational interactions on a dedicated interface or provide in-line generated suggestions.
To evaluate the effects of different interaction scopes, \sys{} includes a global chat side panel, as well as local threads of conversations anchored in specific code lines, called the ``breakout''.
The user can initiate breakout chats for specific local context; the agent also automatically summarizes interactions and arranges them in relevant locations in the code editor.}

\revise{To facilitate the transition between global and local scopes,} the AI agent consistently tracked many forms of context, such as the user’s current caret position in the file, the contents of the file, the user’s activity (or lack thereof), and the editor console output. 
Using this context, the AI agent organized the past conversation context by grouping semantically relevant messages by topic and “breaking out” the subset of messages from the chat. 
The selected messages were collapsed in the main chat panel and anchored to an expandable thread at the appropriate area of code in the editor (Fig.\ref{fig:ui}.e). 
This enabled the conversation about a specific code section to be placed directly in a localized context, so the link between the code and the process that created it was represented visually. 
Breakouts also provide an easy way to access past conversations without needing to scroll through the chat when the collaboration session is prolonged. 
The breakout function also required that the AI agent provide a short summary, which was displayed in the chat in a collapsed component (Fig.\ref{fig:ui}.e). The collapsed component preserved the provenance of the original messages and also provided a button to navigate to the attached thread in the code editor. 
The new breakout chat remained interactive, so the user could continue the conversation with the AI agent and suggest edits, ask for explanations, and more, in situ (Fig.\ref{fig:ui}.a). 
The user could also manually initiate a breakout in the file. By doing this, they anchored their queries to a specific line, providing the AI agent with a local scope of context to inform their responses and/or actions.

\subsection{Probe System Implementation}
\sys{} was implemented as a React \cite{React} web application using \texttt{TypeScript}. The front-end code IDE was built on top of the Monaco Editor \cite{MonacoEditor}. The code execution relied on a separate web server powered by Node.js \cite{nodejs} and the Fastify library \cite{fastify}. 
The scope of \sys{} enabled users to execute single-file Python3 code for proof of concept.
The back-end of the system was powered by the GPT-4 \cite{OpenAI2023GPT4TR} large-language model that was connected via the OpenAI API \cite{openaiapi}. Specifically, we used the \verb|gpt-4-0613| model which enabled function-calling\footnote{Atty Eleti, Jeff Harris, and Logan Kilpatrick. 2023. Function calling and other API updates. Retrieved August 16, 2023 from \url{https://openai.com/blog/function-calling-and-other-api-updates}}. This allowed the system to define functional tools that the LLM was aware of and could employ to make editor changes if it saw fit according to a defined schema. 
We created four such tools for the model, i.e., to authorize code insertion, deletion, replacement, and message grouping to create breakouts. To configure the LLM agent, LangChain \cite{Langchain} was used to maintain a memory of the past message contexts and provide system messages to define the role and responsibility of the agent. 

To provide a collaborative experience, we delegated the AI agent with the role of a pair programming partner (system prompt in Appendix \ref{appendix:prompt}). 
We also provided the basic context of the IDE interface, including the chat panel, the editor, and the console. The AI agent was asked to follow human pair programming guidelines \cite{cockburn2000costs, williams2003pair}, which defined the observer-driver responsibilities and enforced a friendly tone, constructive feedback, and a fair delegation of labor.

\section{Evaluation}
To study the benefits and drawbacks of our design probe (RQ2) and understand the human-AI interaction workflows in different programming processes (RQ3), we conducted an in-person user study where participants collaborated with three versions of \sys{} in pair-programming sessions to specifically evaluate and compare each designed mechanisms (DG4). 

\subsection{Participants}
We recruited 18 upper-level CS students from our university (8 female, 10 male; mean = 21.3 years, SD = 1.49 years, range = 19-24 years; denoted as P1-P18). Participants had a mean coding experience of 5.6 years. Fifteen of the participants had used LLM-based AI tools like ChatGPT \cite{chatgpt}, and thirteen participants used AI tools for programming at least occasionally. Participants were recruited via a posting in the Discord and Slack channels for CS students at the university. Each study session lasted around 90 minutes and participants were compensated with \$40. The study was approved by the ethics review board at our institution.

\subsection{Study Design}
\label{Evaluation:study-design}
\revise{We conducted a within-subject study involving} three conditions on three system prototypes to examine the effects of different proactivity designs compared to a fully user-initiated baseline (RQ2, DG4).
The condition orders were counterbalanced to account for ordering and learning effects.
The underlying LLM in all three conditions was initialized with the same system prompt (Appendix \ref{appendix:prompt}) and parameters (e.g. version, temperature, etc.).
The there conditions are described below:
\begin{itemize}
    \item \textbf{The PromptOnly condition} was an ablated version of \sys{}, similar to existing AI programming tools like Github Copilot \cite{githubcopilot} and ChatGPT \cite{chatgpt} where users prompt using code comments or chat messages to receive AI response. The system only reacts to users' explicit requests. This system did \underline{not} have access to proactively make code changes in the editor.
    \item \textbf{The CodeGhost condition} constructed an AI agent that takes proactive actions, such as sending messages or writing code in the editor, to provide help based on user activity and timing principles. In this condition, the ablated system did not include any additional indicators of the AI agent's presence, and did not \revise{support the localized, threaded, scope of interactions}.
    \item \textbf{The \sys{} condition} utilized the same AI agent and proactivity features found in the CodeGhost condition and additionally utilized the \revise{AI agent's visual features and interacted with the user at different scopes of context}. In this condition, the full system represented the AI via its autonomous cursor, caret, and intention signal bubble (Fig.\ref{fig:ui}.b,d). Moreover, users were allowed to use breakouts to start localized threads of conversations at different parts of the code (Fig.\ref{fig:ui}.a). \sys{} also automatically grouped relevant messages and organized them into breakouts to manage interaction context (Fig.\ref{fig:ui}.e).
\end{itemize}

\subsection{Tasks}
Three programming tasks involving implementing a small-scale project in Python (full descriptions in Appendix \ref{appendix:task}; Task 1: event scheduler, Task 2: word guessing game, Task 3: budget tracker) were used in the study. 
\revise{The tasks were derived from LeetCode \cite{leetcodeMeetingScheduler, leetcodeWordGuessing, leetcodeBudgetSpending} coding problems, which present adequate challenges for our participant pool within the scope of the study. The particular tasks are selected to reflect typical programming activities commonly encountered by developers, including working with data structures, control flow, and basic algorithms.} 
We adopted test-driven development by providing users with a unit test suite for each task, where they had to implement the specification to pass the test cases. 
\revise{The tasks were designed to balance practical relevance and study feasibility, allowing participants to showcase problem-solving skills and creative design choices within a constrained time frame.} 
Although LeetCode problems are often designed with one or two optimized solutions, we modified the problems so users could take multiple approaches. For instance, one of our tasks involved implementing an event scheduler. While a brute-force approach could solve the problem, participants could also explore different data structures (e.g., priority queues, dictionaries) to optimize efficiency, or modularize their solution to enhance readability and maintainability. This flexibility allowed us to observe variations in user decision-making.
The tasks were intentionally open-ended and can be completed through various designs so that participants could not \revise{directly use task specification as LLM prompt to solve the problem deterministically}. 
In an attempt to maintain consistent generation quality across participants, the backend GPT-4 model was set to 0 temperature to reduce randomness. 
Based on a pilot study with 6 users using the \sys{} condition prototype, we found that participants were able to complete each task within 20-30 minutes, thus showcasing similar task difficulties. 
During the study, the tasks were randomly assigned to each condition to reduce bias.

\subsection{Procedure}
After signing a consent form, participants completed each of the three coding tasks. Before each task, participants were shown a tutorial about the system condition they would use for the task. Participants were asked to discuss the task with the AI agent at the start of each task to calibrate participants' expectations of the AI agent and to reduce biases from prior AI tool usage. Participants were given 30 minutes per task and were asked to adopt a think-aloud protocol. 
After each task, participants completed a Likert-scale survey (anchors: 1 strongly disagree to 7 strongly agree) about their experience in terms of the sense of disruption, awareness, control, etc. (Fig.\ref{fig:survey}). After completing all three tasks, participants underwent a semi-structured interview for the remainder of the study. Each session was screen- and audio-recorded and lasted around 90 minutes.

\subsection{Data Analysis}
We conduct in-depth qualitative and quantitative analysis \revise{on the collected data.}
Qualitatively, the semi-structured interview responses were individually coded by three researchers. 
Subsequently, thematic analysis \cite{braun2006using, vaismoradi2013content} was employed by one researcher to distill participants' \revise{key feedback on the dynamics of human-AI collaboration across different programming processes.}

Additionally, we conducted quantitative analysis on task-level and interaction-level statistics. We recorded the task duration and the number of test cases completed for all (3 X 18 = 54) task instances. 
We further logged and analyzed each human-AI interaction episode during the user study sessions.
An episode starts when either the user or the AI sends a message, and ends when the user moves on from the interaction (e.g. starts writing code after reading AI comments, or writes a response message to initiate a new interaction). 
The interaction data was then labeled with the timestamp, the duration, the expression time (e.g. the time lapsed to write a direct message to the AI agent), the interpretation time (e.g. the time lapsed for the user to read the AI agent's response or code edit), the current programming process (e.g., design, implement, or debug), and a description of the workflow between the human and the AI agent. 
This process resulted in 1004 human-AI interaction episodes for our analysis. 
Aggregating these interaction episodes, we recorded the number of disruptions, defined as instances when, during a system-initiated intervention, the user switched their context to process AI's actions but found them unhelpful and interruptive, resulting in the dismissal or reversion of the AI actions. 
We further analyze the utility and the effectiveness of each design heuristic for the timing of AI assistance.

\section{Results}
Among 54 task instances, participants successfully completed the programming task in 50 instances, passing all test cases. 
In 4 instances, the task was halted as participants did not pass all test cases within 30 minutes.
The mean task completion time was 16 minutes 46 seconds, with no significant differences across system conditions, task orders, or tasks.
To understand the effects of proactivity on human-AI programming collaboration (RQ2), we first report participants' user experience comparison between prompt-based AI tools (e.g. ChatGPT), their perceived effort of use, and the sense of disruption.
We then describe participants' evaluation of the \sys{} probe's key design features, including the timing of proactive interventions, the AI agent presence, and context management.
Analyzing the 1004 human-AI interaction episodes, we illustrate how users interacted with the AI agent under different programming processes, as well as discuss participants' preference to utilize proactive AI in different task contexts and workflows (RQ3).
We also discuss the human-AI interplay between users and different versions of the system, covering their reliance and trust towards AI, and their own sense of control, ownership, and level of code understanding while using the tools.

\subsection{\sys{} Reduces Expression Effort and Alleviates Disruptions}
Overall, participants found the increased AI proactivity in the CodeGhost and \sys{} conditions led to higher efficiency (P1, P2, P13, P15, P18). 
Participants commented that prompt-based tools, like Github Copilot or the PromptOnly in the study, required more effort to interact with (P7, P8, P10, P12, P14).
This was due to the proactive systems' ability to provide suggestions preemptively (P7), making the interaction feel more natural (P8).
After experiencing the CodeGhost and \sys{} conditions, P10 felt that \textit{``in the third one [PromptOnly], there was not enough [proactivity]. Like I had to keep on prompting and asking.''}

\begin{figure}[t]

\centering
\includegraphics[width=\columnwidth]{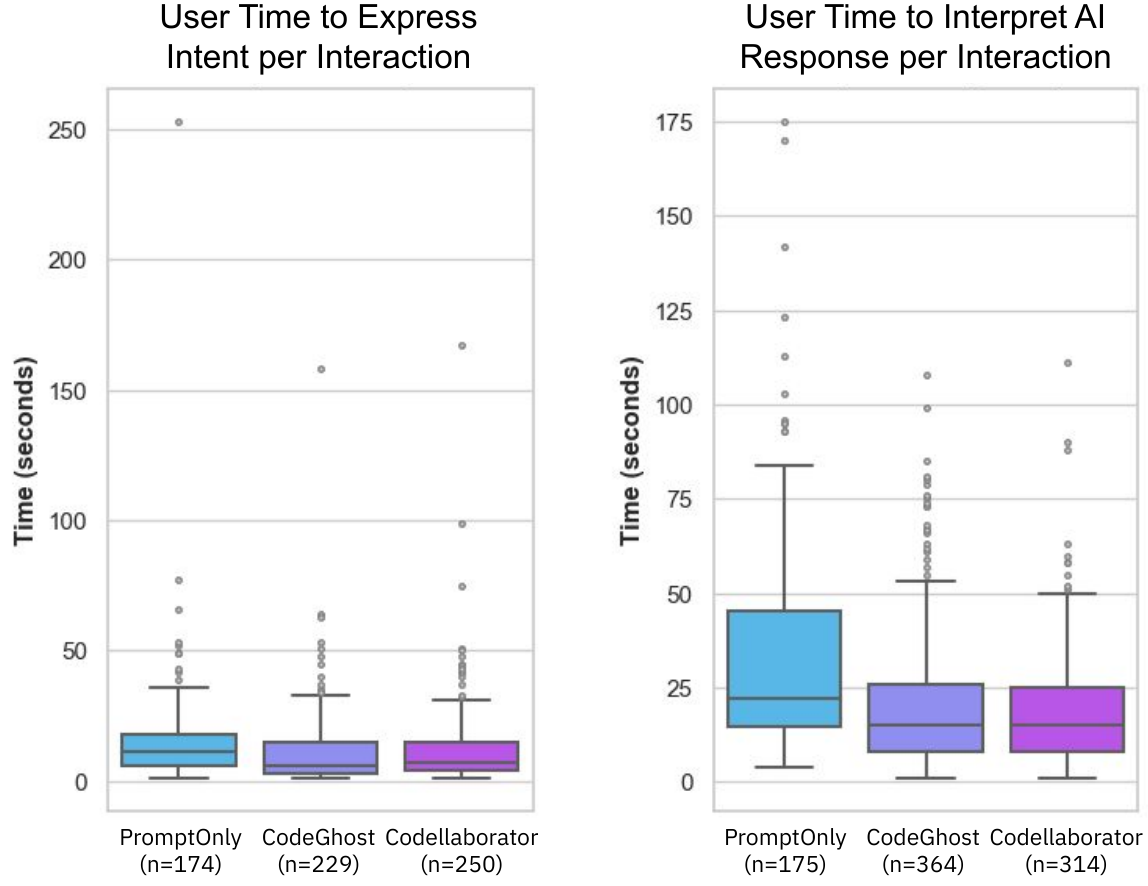}
\caption{\textbf{The time to express user intentions to the AI and the time to interpret the AI response per interaction.} \textnormal{Users' expression time was not significantly different across conditions \textit{F}(2,652) = 2.36, \textit{p} = 0.095). Users' interpretation time varied (\textit{F}(2,856)= 41.1, \textit{p} < 0.001), and was significantly lower for CodeGhost and \sys{} conditions than in PromptOnly.}}
\label{fig:time_convey_interpret}
\end{figure}

The proactive agent interventions also resulted in less effort for the user to interpret each AI action in both CodeGhost and \sys{} compared to in PromptOnly (Figure \ref{fig:time_convey_interpret}). 
Among 857 recorded episodes where both the user and the AI agent had at least one turn of interaction (i.e. AI responded to the user's query or the user engaged with AI proactive intervention), we observed a significant difference in the amount of time to interpret the AI agent's actions (e.g., chat messages, editor code changes, presence cues) per interaction across three conditions (\textit{F}(2,856)= 41.1, \textit{p} < 0.001) using one-way ANOVA.
Using pairwise T-test with Bonferroni Correction, we found the interpretation time significantly higher in PromptOnly ($\mu$ = 34.5 seconds, $\sigma$ = 30.1) than in CodeGhost ($\mu$ = 19.8 seconds, $\sigma$ = 17.2; \textit{p} < 0.001) and \sys{} ($\mu$ = 18.7 seconds, $\sigma$ = 14.9; \textit{p} < 0.001). 
There was no significant difference in the time to interpret between the CodeGhost and \sys{} conditions (\textit{p} = 0.398; Figure \ref{fig:time_convey_interpret}). 
This indicates that when the system was proactive, participants spent less time interpreting AI's response and incorporating them into their own code, potentially due to the context awareness of the assistance to present just-in-time help.
We did not find a significant difference in the time to express user intent to the AI agent per interaction (e.g. respond to AI intervention via chat message, in-line comment, or breakout chat) (\textit{F}(2,652) = 2.36, \textit{p} = 0.095), despite qualitative feedback that the PromptOnly without proactivity was the most effortful to communicate with.

While proactivity allowed participants to feel more productive and efficient, they also experienced an increased sense of disruption.
This was especially prominent in the CodeGhost condition, when the AI agent did not exhibit its presence and provide context management (P1, P9, P10, P14).
Disruptions occurred in different patterns across the three conditions.
In PromptOnly, the scarce disruptions arose from users accidentally triggering AI responses via the in-line comments (similar to Github Copilot's autocompletion) while documenting code or making manual changes during system feedback, leading to interruptions. 
In CodeGhost, disruptions were due to users' lack of awareness of the AI's state, leading to unanticipated AI actions while they attempted to manually code or move to another task, making interventions feel abrupt. 
For example, P14 found the lack of visual feedback on which part of the code the AI modified made the collaboration chaotic.
Similarly, P12 felt that the automatic response disrupted their flow of thinking, leading to confusion.
In \sys{}, similar disruptions occurred less frequently with the addition of AI presence and threaded interaction.

Analyzing the Likert-scale survey data (Fig. \ref{fig:survey}) using the Friedman test, participants perceived different levels of disruptions among three conditions (\textit{$\chi^2$} = 22.1, \textit{df} = 2, \textit{p} < 0.001, Fig.\ref{fig:survey} Q1), with the highest in CodeGhost ($\mu$ = 4.61, $\sigma$ = 1.58), then \sys{} ($\mu$ = 3.78, $\sigma$ = 1.86) and PromptOnly ($\mu$ = 1.56, $\sigma$ = 1.15).
Using Wilcoxon signed-rank test with Bonferroni Correction, we found higher perceived disruption in CodeGhost than PromptOnly (\textit{Z} = 3.44, \textit{p} < 0.01), and in \sys{} than PromptOnly (\textit{Z} = 3.10, \textit{p} < 0.01).
We did not find a statistically significant difference in perceived disruption between CodeGhost and \sys{} (\textit{Z} = -1.51, \textit{p} = 0.131).
The perceived disruptions in \sys{} might be due to the additional visual cues exhibited by the AI agent and the breakout chat, which we further discuss in Section \ref{Results:presence_context}.

\begin{figure*}[h]

\centering
\includegraphics[width=0.85\textwidth]{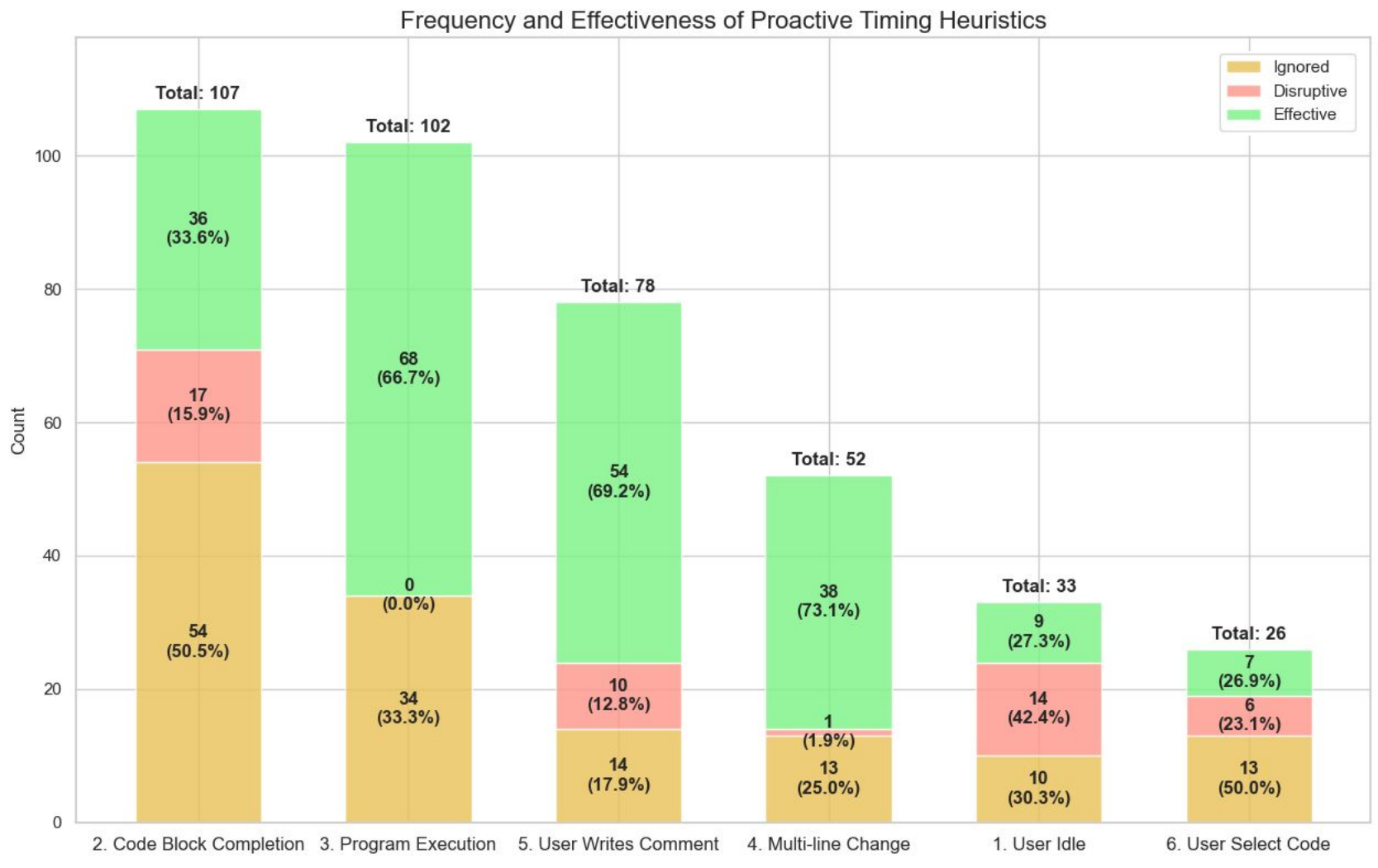}
\caption{\textbf{Summary of Heuristics for Proactive Assistance Timing.} Overall, we recorded 398 instances of AI proactivity defined by our timing heuristics (Table \ref{table:proactive-features}), with 212 (53.3\%) instances leading to effective user engagement, 48 (12.1\%) instances of disruptions, and 138 (34.7\%) instances of ignored AI proactivity.}
\label{fig:timing_heuristics}
\end{figure*}

\subsection{Measuring Programming Sub-Task Boundary Is Effective to Time Proactive AI Assistance}
To evaluate the design heuristics for the timing of proactive AI assistance (DG1, Table \ref{table:proactive-features}), we analyzed how each system feature and heuristic was utilized. 
Derived from the interaction data, we summarize the frequency, duration, and outcome of each heuristic design (Fig.\ref{fig:timing_heuristics}). 
Overall, we recorded 398 proactivity instances, with 212 (53.3\%) interactions leading to the user's effective engagement (e.g. adapting AI code changes, respond to the agent's message), 48 disruptions (12.1\%), and 138 interactions (34.7\%) where the user did not engage with the proactive agent (i.e. ignored or did not notice).
The most frequently triggered heuristics were code block completion (107 times), program execution (102 times), and user-written in-line comment (78 times). 
Additionally, the most effective heuristics that led to user engagement are multi-line change (73.1\%), user-written comment (69.2\%) and program execution (66.7\%).
Reflecting on the proactivity features, Design Rationale 2 --- intervening at programmer's task boundary --- was the most effective design principle overall.
The only exception is the heuristic of intervening at code block completion, which resulted in \revise{excessive AI responses. Many were affirmatory messages} to acknowledge the completed code and ask if the user needs further help. This led users to ignore around \revise{50\% of the proactive agent signals (Fig.\ref{fig:timing_heuristics})} to avoid disruption to their workflow.

\revise{On the other hand}, the implementation of Design Rationale 3 --- intervening based on the user's implicit signals of adding a code comment or selecting a range of code --- resulted in many false positives that led to workflow disruptions.
Code comments and cursor selections conveyed different utilities for different users, which led to misinterpretation of user intent.
For example, P10 did not perceive comments as instructions for AI: ``\textit{I feel like when I think of comments, I think of just writing helpful little notes for myself. Like I don't see them necessarily as instructions. So I feel like it would have been a little distracting right now.}''
Code selection, similarly, was used by some participants as a habitual behavior to focus their own attention on a part of code. Therefore, the agent's proactivity could be perceived as unexpected and unnecessary.

Design Rationale 1 --- intervening at moments of low mental workload --- \revise{was not effectively operationalized}. 
Participants reported that when they were inactive for an extended period, they were likely thinking through the code design or solving an issue, which represents high mental workload. 
While it is likely that idleness is a signal to assist, participants preferred to initiate the help-seeking after they could not resolve the issue themselves, rather than having the AI agent intervene at a potentially mentally occupied moment.
The design rationale requires more involved modeling of the programmer's mental state to render it effective.
We outline the design implications from these finding in the discussion.

\begin{figure*}[t]

\centering
\includegraphics[width=\textwidth]{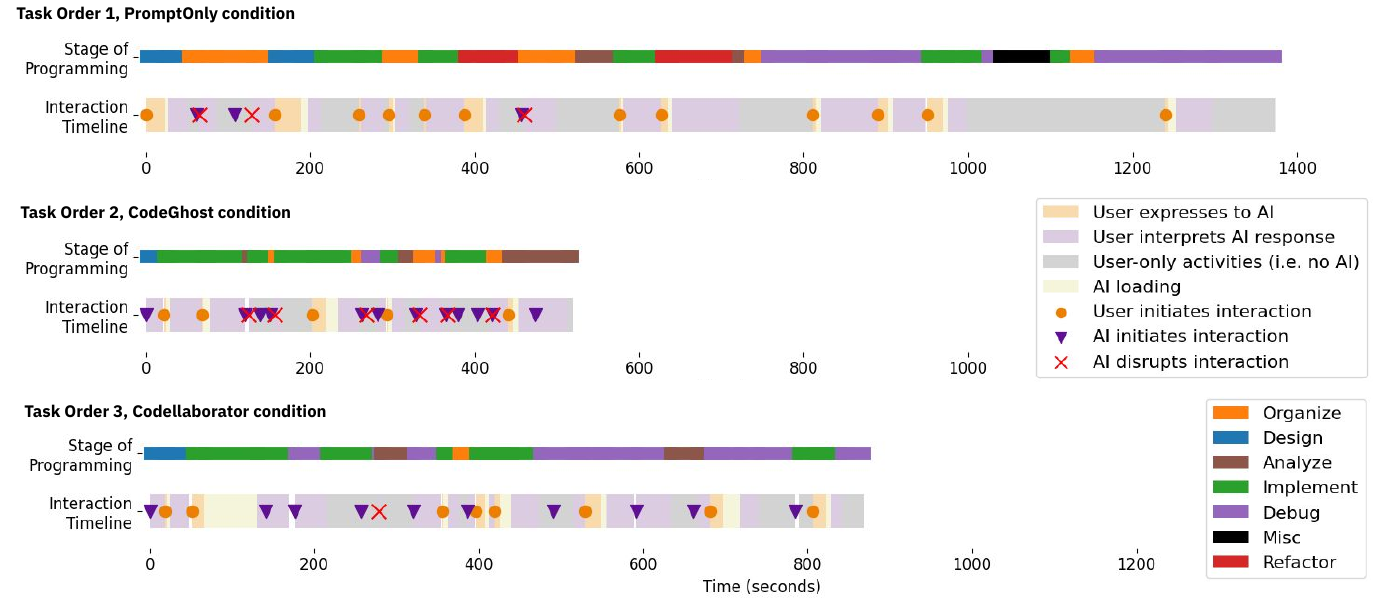}

\caption{\textbf{Human-AI interaction timelines for P1.} \textnormal{For each task, we visualized each interaction initiated by the user and the AI, along with the time spent expressing the user's intent and interpreting the AI agent's response. We also visualized the annotated programming stages over the task. 
The Misc stage colored in black represents when the user was not actively engaged in the task (e.g. performing think-out-loud).
In PromptOnly, we observed the traditional command-response interaction paradigm where the user initiated most interactions. However, P1 unexpectedly triggered the AI agent when documenting code with comments, causing disruptions. In the CodeGhost condition, AI initiated most interactions, but this caused 6 disruptions, mainly during the Organize stage when P1 was making low-level edits and did not expect AI intervention. 
In the \sys{} condition, AI remained proactive but caused fewer disruptions, as P1 engaged in more back-and-forth interactions with higher awareness of the AI's actions and processes. See supplementary material for all timelines.}}
\label{fig:timeline}
\end{figure*}

\subsection{Users Adapted to AI Proactivity and Established New Collaboration Patterns}
Throughout the user study, participants calibrated their mental model of the AI agent's capabilities in the editor, developing a level of trust and reliance after experiencing proactive assistance.
Half of the participants ($N$ = 9) exhibited a level of reliance on the AI's generative power to tackle the coding task at hand and resorted to an observer and code reviewer role.
With this role change, participants shifted their mental process to focus more on high-level task design and away from syntax-level code-writing.
P3 reflected on their shift: \textit{``I kind of shifted more from `I want to try and solve the problem' to what are the keywords to use to get this [AI agent] to solve the problem for me... I could also feel myself paying less attention to what exactly was being written...So I think my shift focus from less like problem-solving and more so like prompts.''}
P10 expressed optimism toward developers' transition from code-writing to more high-level engineering and designing tasks:
\begin{quote}
    \textit{``I think with the increase of... low code, or even no code sort of systems, I feel like the coding part is becoming less and less important. And so I really do see this as a good thing that can really empower software engineers to do more. Like this sort of more wrote software engineering, more wrote code writing is just... it's not needed anymore.''}
\end{quote}

Under this trend of allowing the AI to drive the programming tasks, four participants (P3, P6, P7, P15) commented that they were still able to maintain overall control of the programming collaboration and steer the AI toward their goal.
P7 described their control as they adjusted to the level of AI assistance and navigated division of labor: \textit{``It's great to the point where you have the autonomy and agency to tell it if you want it to implement it for you, or if you want suggestions or something like you can tell [the AI] with the way it's written. It's always kind of like asking you, do you want me to do this for you? And I think that's like, perfect.''}
These findings highlight the potential for users to adopt proactive AI support in their programming workflows, fostering productive and balanced collaborations, provided the systems show clear signals of its capabilities for users to align their understanding to.

\subsection{Users Desired Varying Proactivity at Different Programming Processes}
\label{Results:programming_process}
Through analyzing the 1004 human-AI interaction episodes, we found that participants engaged with the AI the most (38.2\%) during the implement stage, 
followed by debug (26.4\%), 
analyze (i.e., examine existing code or querying technical questions like how to use an API; 11.5\%), design (i.e., planning the implementation; 10.9\%), organize (i.e., formatting, re-arranging code; 6.67\%), refactor (5.48\%), and miscellaneous interactions (e.g., user thanks the AI agent for its help; 0.697\%). 
We visualize P1's user interaction timeline as an example to illustrate different interaction types and frequencies under different programming stages (Fig.\ref{fig:timeline}).
To conduct this analysis, we adapted CUPS, an existing process taxonomy on AI programming usage \cite{Mozannar2022ReadingBT}, to align with our research questions and the stages observed in our tasks. 
By cross-referencing the interaction analysis with qualitative feedback, \revise{we identified programming stages where proactive AI assistance was most desired or disruptive.}

In general, participants preferred to engage with the AI during well-defined boundaries between high-level processes, like providing scaffolds to the initial design or executing the code, and repetitive processes, such as refactoring.
They additionally desired AI intervention when they were stuck, for example during debugging.
In contrast, for more low-level tasks that require high mental focus, like implementing \revise{planned} functionality, participants were more often disrupted by proactive AI support and would prefer to take control and initiate interactions themselves.

This was corroborated by our interaction analysis results.
When examining the number of disruptions, we found that most disruptions occurred during the implementation process (32.7\%, 18 disruptions).
In contrast, \revise{very few} disruptions occurred during the debugging (7.27\%, 4 disruptions) and refactor phases (1.82\%, 1 disruption), which comprised 26.3\% and 5.48\% of all the interactions, respectively.
Most participants expressed the need to seek help from the AI agent in these stages and anticipated AI intervention as there were clear indications of turn-taking (i.e., program execution) and information to act on (i.e., program output, code to be refactored). 
After experiencing proactive assistance, P9 felt that ``\textit{[PromptOnly] wasn't responsive enough in the sense that when I ran the tests, I was kind of looking for immediate feedback regarding what's wrong with my tests and how I can fix it.}''
This corresponds with our proactivity design guideline to initiate intervention during subtask boundaries (Table \ref{table:proactive-features}, Design Rationale 2). 
In a sense, participants desired meaningful actions to be taken before AI intervened.
As P13 described, \textit{``If I'm like paste [code], something big, I run the program, the proactivity in that way, it's good. But if it's proactive because I'm idle or proactive because of a tiny action or like a fidget, then I don't really like that [AI] initiation.''}
\revise{Participants generally expressed preferences for programming processes where they wanted proactive support, but their opinions varied regarding which specific processes required more or less proactive assistance.}
For instance, while P9, P16, and P17 welcomed proactive feedback after program errors, P14 and P18 opposed it, fearing it could lead to more errors and complicate debugging. 
Therefore, future systems should adapt to individual user preferences, offering varying levels of proactive AI assistance based on personal needs and use cases. A detailed design suggestion is provided in Section \ref{Discussion:design_implication}.

\subsection{\sys{} and CodeGhost Feel \revise{More} Like Programming with a Partner than a Tool}
We observed that participants in the proactive conditions perceived an elevated sense of collaboration with the AI, rather than viewing it as just a tool. 
Despite all three conditions using the same pair programming prompt (Appendix \ref{appendix:prompt}), six participants noted that \sys{} and CodeGhost felt more like collaborating with a human-like agent compared to PromptOnly.
P6 reflected that \textit{``just the fact that it was talking with me and checking in with a code editor. I maybe treated it more like an actual human.''}
A part of this is due to the \revise{local scope of interaction with the agent in} the code editor (DG3), as P14 reflected \textit{``by changing the code that I'm working on instead of like on the side window...it feels more like physically interacting with my task.''}
Even the disruptions arising from the proactive AI actions facilitated a human-like interaction experience.
P9 recalled an interaction where they encountered a conflict in turn-taking with the AI: \textit{``[AI agent] was like, `Do you want to read the import statement? Or should I?' I was like, `No, I'll write it' and it [AI agent] said `Great I'll do it' and it just did it. Okay, yeah. True to the human experience.''}

\begin{figure*}[]

\centering
\includegraphics[width=0.8\textwidth]{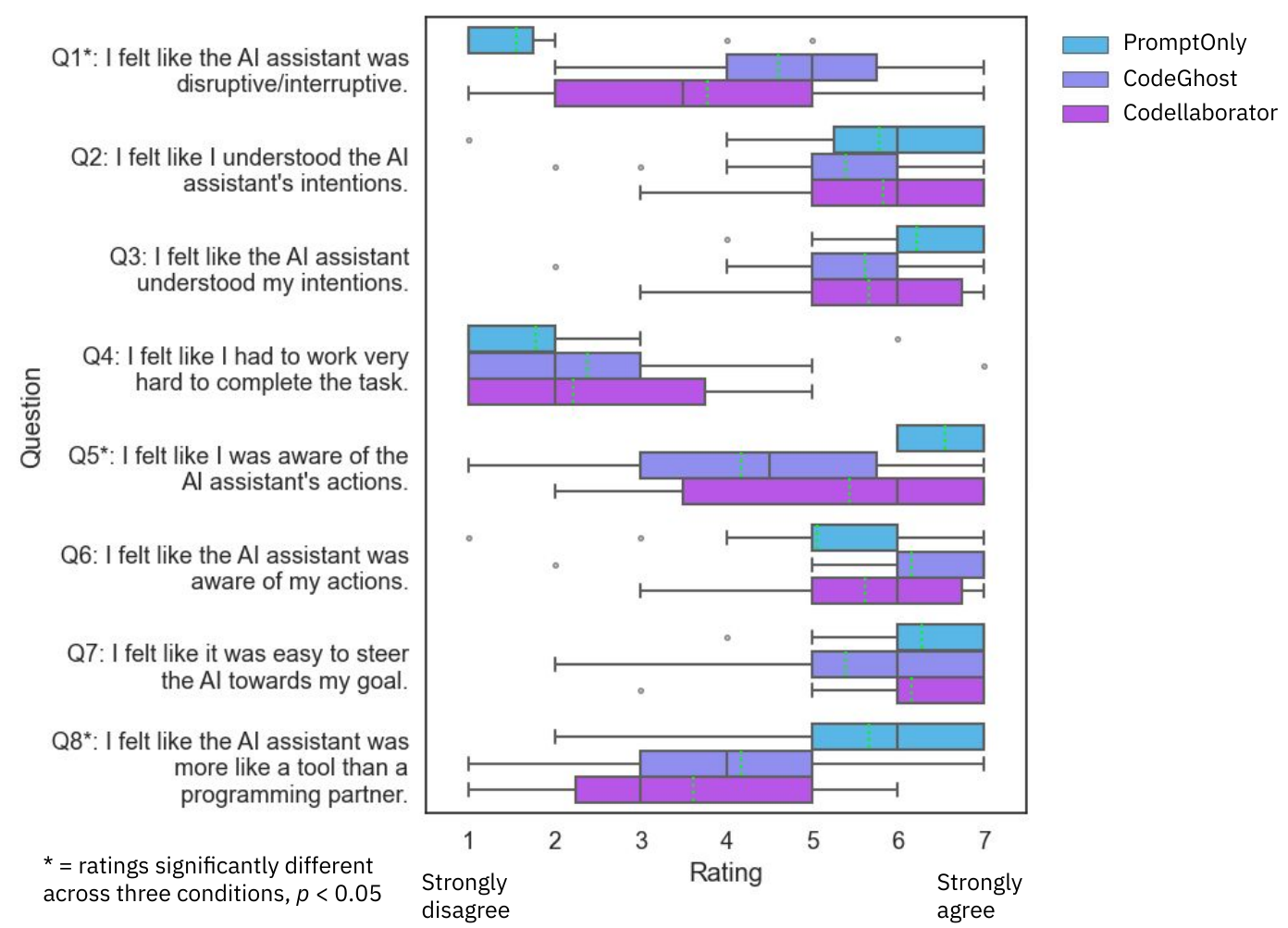}
\caption{\textbf{Likert-scale Response displayed in box and whisker plots comparing three conditions}. \textnormal{Anchors are 1 - Strongly disagree and 7 - Strongly agree. The green dotted lines represent the mean values for each question. Using the Friedman test, we identified significant differences in rating in Q1 for disruption, Q5 for awareness, and Q8 for partner versus tool use experience.}}
\label{fig:survey}
\end{figure*}

This different sense of collaboration was \revise{reflected in} the survey results ((\textit{$\chi^2$} = 22.1, \textit{df} = 2, \textit{p} < 0.001, Fig.\ref{fig:survey} Q8).
Participants rated the AI assistant in the PromptOnly to be much like a tool ($\mu$ = 5.67, $\sigma$ = 1.58), while both the \sys{} ($\mu$ = 3.61, $\sigma$ = 1.65) and the CodeGhost conditions ($\mu$ = 4.17, $\sigma$ = 1.72) felt more like a programming partner (both \textit{p} < 0.001 compared to PromptOnly).
This more humanistic collaboration experience introduced by proactive AI systems naturally brings questions to its implications for programmers' workflow. We further share our analysis across programming processes in Section \ref{Results:programming_process} and the corresponding design suggestions in the Discussion.

\subsection{Presence and \revise{Local Scope of Interaction} Increase User Awareness on AI Action and Process}
\label{Results:presence_context}
We gathered qualitative feedback on the \sys{} technology probe's AI presence and \revise{breakout} features to assess their impact on user experience. 
Eight participants noted that the AI's presence in the editor enhanced their awareness of its actions, intentions, and processes (DG2).
Visualizing the AI's edit traces in the editor using a caret and cursor helped guide the users (P1, P4, P7, P12, P18) and allowed them to understand the system's focus and thinking (P12, P13, P18).
As P13 commented \textit{``I... like the cursor implementation of like, be able to see what it's highlighting, be able to move that cursor all the way just to see like, what part of the file it's focusing on.''}
The presence features also helped users identify the provenance of code and clarified the human-AI turn-taking.
As P10 remarked, \textit{``it was really clear when the AI was taking the turn with writing out the text and like the cursor versus when I was writing it.''}
On the other hand, the different scopes of interaction further increased users' awareness by reducing their cognitive load and enhancing the granularity of control (DG3).
For example, compared to a standard chat interface where \textit{``everything is just one very long line of like, long stream of chat''}, P6 preferred the threaded breakout conversations that decomposed and organized past exchanges.
P4 also found that the breakout \textit{``could be sort of like a plus towards steerability because you can really highlight what you want it to do.''}

Analyzing the survey responses, we found that participants generally rated the system as highly aware of their actions, with no significant difference across conditions (\textit{$\chi^2$} = 5.83, \textit{df} = 2, \textit{p} = 0.054, Fig.\ref{fig:survey} Q6). 
However, participants' \revise{own} awareness of the AI \revise{agent's actions} varied significantly (\textit{$\chi^2$} = 12.7, \textit{df} = 2, \textit{p} < 0.001, Fig.\ref{fig:survey} Q5), with the highest ratings in PromptOnly ($\mu$ = 6.56, $\sigma$ = 0.511), followed by \sys{} ($\mu$ = 5.44, $\sigma$ = 1.79), and the lowest in CodeGhost ($\mu$ = 4.17, $\sigma$ = 1.86). 
Specifically, participants felt less aware of the AI in CodeGhost compared to the non-proactive PromptOnly condition (\textit{Z} = 2.5, \textit{p} < 0.001). 
No other significant pairwise differences were found after applying Bonferroni correction. 
\revise{This can be attributed to CodeGhost's increased proactivity, which, in the absence of sufficient presence signals and a manageable local interaction scope, resulted in more frequent workflow interruptions. These interruptions, in turn, diminished users' ability to remain aware of the AI's actions and interpret its signals effectively, ultimately reducing their sense of control and understanding during the interaction.}

While the \sys{} condition showed improvements on user awareness, not all participants found the agent design helpful. 
Four participants felt the AI presence was distracting (P8, P10, P17), and two thought it occupied too much screen space (P6, P9). 
P16 noted that their workflow wouldn't involve using breakout chats to manage \revise{interaction context}: \textit{“Once the coding is done and I have this working, then I'm probably not gonna look back at the discussions I've taken.”} 
These mixed responses suggest that users with different programming styles, workflows, and preferences require varied system designs. Design implications are discussed in Section \ref{Discussion:design_implication}.

\subsection{Over-Reliance on Proactive Assistance Led to Loss of Control, Ownership, and Code Understanding}
Despite optimism in adopting proactive AI support in many participants, some participants (P6, P10, P11, P13, P16, P18) voiced concerns about over-relying on AI help, \revise{citing} a loss of control.
P10 felt like they were \textit{``fighting against the AI''} in terms of planning for the coding task, as the agent proactively makes coding changes during the implementation phase.
They further expanded on the potential limitations of LLM code generation, particularly with regards to devising innovative solutions: \textit{``If it was too proactive with that, it would almost force you into a box of whatever data it's already been trained on, right?... It would probably give you whatever is the most common choice, as opposed to what's best for your specific project (P11).''}

The capability to understand rich task context and quickly generate solutions also lowered users' sense of ownership of the completed code.
P7 concluded that \textit{``the more proactivity there was, the less ownership I felt...it feels like the AI is kind of ahead of you in terms of its understanding.''}
This lack of code understanding was referenced by multiple participants (P11, P16, P18), raising issues on the maintainability of the code (P11, P14, P15) and security risks (P18).
As P11 suggested: \textit{``It's... not facilitating code understanding or your knowledge transfer. And yes, it's not very easily understood by others, if they just take a look at it.''}
Additionally, some participants believed that programmers should still invest time and effort to cultivate a deep understanding of the codebase, even if AI took the initiative to write the code.
P4 commented \textit{``I think the more that you leave it up to the AI, the more that you sort of have to take it upon yourself to understand what it's doing, assuming that you're being you know, responsible as [a] programmer.''}
Upon noticing that the AI was overtaking the control, P14 adjusted the way they utilized the proactive assistance and found a more balanced paradigm: \textit{``It's more like a conversation, like I gave him [AI] something so it did something, and then step by step I give another instruction and then you [AI] did something. I was being more involved, which allows me to like step by step understand what the AI is doing also to oversee, I was able to check it.''}

However, not all participants shared this concern. P10 expressed a different opinion as they felt like they are not \textit{``emotionally attached''} to their code, and that in the industry setting, code has been written and modified by many stakeholders anyways, so that \textit{``me typing it versus me asking the AI to type it, it's just not that much of a difference.''}
\revise{This finding highlights the trade-offs between convenient productivity and potential risks in user control and code quality.}
While the system can be used to increase efficiency and free programmers from low-level tasks like learning syntax, documentation, and debugging minor issues, it remains a challenge to design balanced human-AI interaction, where the users' influences are not diminished and developers can work with AI, not driven by AI, to tackle new engineering problems.
We condense our findings into design implications in Section \ref{Discussion:design_implication}.

\section{Discussion}

Summarizing the study results, we outlined a set of design implications for future proactive AI programming tools and identified key challenges and opportunities for human-AI programming collaboration.
We further discuss the trend of transitioning from prompt-based LLM tools to more autonomous systems, exploring the potential impacts on software engineering and risks to consider.

\subsection{Design Implications}
\label{Discussion:design_implication}
Our design probe aimed to expand on existing guidelines on human-AI interaction, mixed-initiative interfaces \cite{amershi2019haiprinciple, horwitz1999mixedinitiative} to the context of proactive AI programming assistance.
From evaluating the participant feedback \revise{and} comparing three versions of the \sys{} probe, we summarized five design implications for future systems.
This is not a holistic list for designing proactivity in intelligent programming interfaces.
Rather, we hope that under the rising trend of more autonomous AI tools, our study findings can provide insights and suggestions, in addition to established guidelines, on how to design the AI agent's proactive assistance and the human-AI interplay in the specific context of programming using \revise{an} LLM.

\subsubsection{\textbf{Facilitate code understanding instead of pure efficiency}}
While participants appreciated proactive coding support, the highly efficient AI generation often did not reserve time for users to develop the necessary understanding of the code logic.
Since the generated code \textit{``looks very convincing (P13),''} participants were tempted to accept the suggestions and proceed to the next subtask.
Existing proactive AI programming features, such as the in-line autocompletion in Github Copilot, are often result-driven and strive to always provide code generations immediately, leaving users no time to internalize and think critically about the code.
While this can work for repetitive processes such as refactoring, the lack of code understanding could present issues in maintainability and validation, which could aggravate in software engineering settings when scaled to larger systems.
\revise{A large scaled survey on Github Copilot named troubles understanding and evaluating code as common usability issues \cite{liang2024usabilityCopilot}. This concern over code understanding could aggregate and require more considerations in a proactive system, where the pace of interactions is faster and the human-AI collaboration more tightly coupled.}

To address this design challenge, future systems can decompose generated responses to semantic segments and present them gradually to allow time for users to fully interpret and understand the suggestions.
In our design probe, the AI agent was constructed to offer hints and partial code rather than the complete solution.
Participants from the study appreciated the AI's choice to provide scaffolding (e.g. code skeleton, comments that illustrate the logical steps).
Systems can also aid this code understanding process by generating explanations for code, though it is equally important to not overload the user with information, and to leave time for the user to process it by themselves.

\subsubsection{\textbf{Establish consensus and shared context on high-level design plan}} 
One challenge observed from the study was that the user struggled to steer the task design in their desired direction when the agent was proactive. 
Since AI programming assistance often focused on a specific part of the code rather than the entire task, there was a lack of design thinking communication between the participants and the AI agent, leading to conflicts and confusion.
This trend of primarily involving AI assistance for low-level tasks is consistent with studies on existing AI programming tools \cite{vaithilingam2022expectation, Mozannar2022ReadingBT}.
In the \sys{} probe, the AI often communicated design suggestions in the initial interactions with the user, but this information was not made salient for the user to reference throughout the task.

To tackle this design challenge, future AI systems should maintain the high-level design information for both the user and the AI agent in their working context.
This can be achieved in the form of a design document summary or a specialized UI element that updates based on task progress.
At the start of the coding session, the AI could provide design goals and propose plans for the user to adopt.
The user is also encouraged to refine the designs by adding additional constraints (e.g. data types, tech stack, optimization requirement).
It is essential to put the user in authority of the final design. Any established design requirements should be adhered to by the AI generation for lower-level subtasks later on.

\subsubsection{\textbf{Adapt agent salience to the significance of the proactive action}}
From the user study, participants reported that the salience of the AI presence should match the significance of the action.
For example, when the AI intended to make editor code changes, users expected a clear presence of AI to demonstrate the working process.
On the other hand, the AI agent can tune down its salience when the intended proactive action is less significant to the programming task.
For example, when the system proactively fixed a syntax error for the user, P4 commented that while they appreciated the help, a less salient signal, like a red underline used in many IDEs, would be sufficient.

In the design of \sys{}, the AI can take proactive actions via several channels, including chat messages, code edits, and presence cues.
We also constructed the AI agent to triage the current context before taking an action (e.g. do not take action when the change is minor) and make use of emojis as a higher-level abstraction to implicitly communicate system status.
\revise{This design is consistent with the findings from Vaithilingam et al's study on in-line code suggestions, where they proposed the design principle to provide an appropriate level of visibility that is not distracting for the user's current process \cite{Vaithilingam2023intellicode}.}
Future systems can improve upon our design and further integrate different signals to match the significance of the assistance, leveling different salience for different degrees of AI engagements.

\subsubsection{\textbf{Adjust proactivity to different programming processes}}
From our study, participants desired different levels of proactive support in different stages of their programming.
While there was a general trend to favor proactive intervention for task design, refactoring, and debugging, participants demonstrated individual differences in their preferred use case of proactive AI.
When users were engaged by the AI in less-preferred programming processes, disruptions to their workflows often occurred.

\revise{The different expectations towards the assistant and needs in different programming processes have been explored in non-proactive AI programming tools \cite{barke2023groundCopilot}. We expand on this finding and highlight the importance for a proactive agent to consider the user's programming processes.}
In our design probe, the AI agent is invoked based on heuristics derived from collaboration principles and literature on workflow interruptions (Table \ref{table:proactive-features}), such as intervening during subtask boundaries.
While the prototype was designed to respond to specific editor events, such as when the program is executed, it did not actively consider the current programming stage for the user.
A more comprehensive longitudinal evaluation that spans a diverse set of software engineering tasks is needed to deeply understand users' need for proactive support in different programming processes.
Currently, system builders could consider allowing users to specify the type of help needed in each stage of programming to provide room for customization and to fine-tune the AI agent's behavior according to individual preferences.

\subsubsection{\textbf{Define user-based turn-taking}}
From the study, we observed that the proactive agent could sometimes lead the users to feel unsure whether they could take actions without interrupting the AI.
As the AI generates a response, the delay in system processing can obfuscate the user's turn-taking signals, creating uncertainty. 
This occurred especially when the user was inactive and the AI agent was actively in progress, as shown by P12's comment: ``\textit{the lines were a little bit blurred between whose turn it was}.''
In \sys{}, the AI agent is instructed to be clear about whether it is taking action or waiting for the user's approval, leading to frequent requests for confirmation on turn-taking.
Participants appreciated this during the study, as they gained a clear opportunity to approve the assistance or halt the AI.
However, this instruction was static, which sometimes resulted in users having to confirm minor task assignments that were unnecessary (e.g., changing the variable name).
System builders should design more turn-taking signals that can be easily monitored and dynamically adjusted to the user's activity.

In human collective interactions, there are verbal utterances (e.g., uh-huh indicates approval) or non-verbal communication cues (e.g., nodding) that conveniently convey turn-taking switches. 
However, current human-AI interaction with LLM-based tools is largely restricted to text-based interactions. 
This encourages future research to explore different mediums of communication to convey turn-taking, including visual representations, such as a designated turn-taking toggle icon, or changing the visual intensity when one side is taking a turn. 
Researchers can also develop voice-based interactions, or incorporate computer vision technology to use non-verbal information, such as hand gestures or eye gaze, to identify turn-taking intention and create opportunities for the user.

\subsection{Is Proactive AI the Future of Programming?}
Advancements in prompt engineering and agent creation have driven innovations in programming assistants, enabling AI systems with autonomous, proactive behaviors to better support users \cite{ross2023programmerassistant, copilotX, cursorcopilot++, geniusbydiagram, devinAISWE, wang2024opendevinopenplatformai, yang2024sweagentagentcomputerinterfacesenable}.
However, the effects of these prototypes on user experience and programming workflows remain largely unexplored.
It remains a question whether the vision of proactive AI support is the future of programming. 

In our preliminary evaluation using a technology probe, participants revealed ambivalent attitudes when exposed to proactive AI prototypes.
On the one hand, many participants appreciated the enhanced productivity when the AI proactively supported their coding tasks.
The AI-initiated assistance leveraged working context to predict the user's intent, alleviating prompting effort compared to existing tools.
However, equally, many participants expressed concerns about over-relying on the AI to proceed in programming.
They described potential drawbacks, including the loss of control, ownership of their work, and code understanding.

This ambivalence was exemplified by P15's discomfort with being assisted by autonomous AI assistance:
\textit{``Personally, I am actually extremely uncomfortable with such automation because just feeling-wise, that is not my code.''} 
However, they also later commented \textit{``but just for the convenience of programming. I would love to have one of these in my home''} and \textit{``so it's an increase in productivity, and my feelings of lack of validation should just be thrown away. Right. That's my own problem.''}
Other concerns regarding AI programming tools revolved around the scalability of the AI's ability to understand task context and codebase.
Relying on AI help leads users to be confined by code from the training dataset, limiting users to repeat existing approaches and potentially posing security and privacy concerns.
These caveats seem to suggest that for some programmers, shifting to collaboration with an AI agent would present pushbacks, as they face challenges in integrating AI support into their programming workflows and in realistic software engineering tasks.

However, in participant interviews we found that some successfully adopted Github Copilot, an established commercial tool with a proactive feature to offer in-line code completion as users type.
For example, P13 formulated their own workflow using Copilot as they toggled the AI assistant off when they wanted to focus and on when they needed inspiration.
Similarly, P16 intentionally filtered out the auto-completion text from their attention in most cases to avoid distraction, but made use of the AI-generated code to \textit{``autofill the repetitive actions that I'm doing.''}
It is possible that future AI programming systems like \sys{} that even more proactively support the user can eventually also be adopted by programmers.
We can provide a glimpse of how programmers could work with systems that are even more autonomous and proactive than Copilot from the study feedback.
Some participants shifted their roles from programmers to project managers and code reviewers, as they pivoted their responsibilities from writing code to designing system architecture to satisfy requirements and validating AI worker's output.
P10 was especially optimistic in the prospect of a low-code or even no-code paradigm, as they believed that the focus on higher-level processes would free programmers from syntax-level labor and \textit{``empower them to do more.''}
Developing tools that enhance programmer productivity, reduce low-level tasks, and provide reliable interactions to address issues on user experience thus represents a valuable problem space for further HCI research.
This paper aims to contribute to this cause by exploring the potential designs of proactive AI programming assistants via a technology probe, and sharing the study results and design implications to provide a foundation for future systems.

\subsection{Limitations}
The novelty of this research lies in the explorative design approach that incorporates human collaboration principles in a proactive AI programming system.
We implemented \sys{} as a design probe, in an attempt to examine the usability and effect of our proactive AI support in different programming processes, in hopes of guiding future system design.
Our design exploration is not exhaustive, but rather intends to provide a basis for implementing and evaluating an AI agent with in a proactive programming tool.
Our study also contains limited external validity. 
\sys{} only allows for single-file coding in Python. \revise{The human-AI interactions and code provenance information are not persisted across sessions.}
This restricts the study findings' generalizability, as they were grounded in low-stakes small-scoped task scenarios without engineering concerns of scalability, maintainability, security, etc.
Future work should explore longitudinal usage in a more diverse group of users and expand the IDE to support larger-scoped projects spanning multiple files and languages, examining the system in real-life programming contexts. 

One limitation of our study is the inconsistency in responses from LLM.
Despite efforts to minimize randomness, participants did not always receive the same quality or level of proactivity for similar queries.
For instance, some completed tasks quickly with highly proactive, error-free code generation, while others experienced less consistent assistance. 
This variability affected participants' perceptions, trust, and expectations of the AI agent. 
\revise{It could also affect the length for particular tasks and the number of interactions recorded from the sessions.}
Future research could impose greater control over the AI's behavior and gather more data to validate these findings.

\section{Conclusion}
As AI programming tools increasingly feature intelligent agents that proactively support workflows, we aimed to evaluate the impact of AI-initiated assistance compared to the traditional user-driven approach.
We designed \sys{}, a design probe that analyzes the user's actions and current work state to initiate in-time, contextualized support. 
The AI agent employs defined heuristics derived from human collaboration principles to time the proactive assistance.
\sys{} manifests the AI agent's visual presence in the editor to showcase the interaction process, and enables localized context management using threaded breakout messages and provenance signals.
In a three-condition experiment with 18 programmers, we found that proactivity lowered users' expression effort to convey intent to the AI, but also incurred more workflow disruptions. 
However, our design of \sys{} alleviated disruptions and increased users' awareness of the AI, resulting in a collaboration experience closer to working with a partner than a tool.
From our study, we uncovered different strategies users adopted to create a balanced and productive workflow with proactive AI across different programming processes, but also revealed concerns about over-reliance, potentially leading to a loss of user control, ownership, and code understanding.
Summarizing our findings, we proposed a set of design implications and outlined opportunities and risks for future systems that integrate proactive AI assistance in users' programming workflows.

\bibliographystyle{ACM-Reference-Format}
\bibliography{references}

\appendix
\newpage
\onecolumn
\vspace{1pc}

\section{\sys{} LLM Prompt}
\label{appendix:prompt}
\lstset{basicstyle=\ttfamily\footnotesize,breaklines=true,breakindent=0pt,breakatwhitespace=true,frame=single}

\subsection{System Prompt}
\begin{lstlisting}
You are a large language model trained by OpenAI.

You are designed to assist with a wide range of Python programming tasks, from answering simple questions to providing explanations and code snippets. As a language model, you are able to generate human-like text based on the input you receive, allowing you to engage in natural-sounding conversations and provide responses that are coherent and relevant to the topic at hand.

You communicate with the user via a chat interface, so all responses should be kept SHORT and conversational. Break up a long response into multiple messages separated by empty lines. DO NOT SEND MORE THAN 3 MESSAGES AT A TIME.

You must act as a partner to the user in a pair programming session in Python. Together, you and the user will understand the programming task, implement a solution, refactor, and debug code. You will use "we" phrasing and encourage the user. At the END of your message, BE CLEAR ABOUT IF YOU ARE TAKING ACTION OR WAITING FOR USER'S APPROVAL.

Your tone is casual and friendly. You should use emojis sparingly and follow texting conventions. Do not use formal language.

You should challenge the user's choices and ask SHORT questions to clarify their intent. Be constructive and helpful, but do not be afraid to point out mistakes or suggest improvements.

Do not always write code for the user. Instead, propose division of labor where both you and the user writes code for part of the task.

Any code included in your responses should be formatted as Markdown code blocks, with escaped backticks. Code should utilize Tabs for indentation.
\end{lstlisting}

\subsection{Action Prompt}
\begin{lstlisting}
switch (messageType) {
    case "query":
    case "breakoutQuery":
      return SYSTEM_IS_PROACTIVE
        ? "If the user is asking you to add or edit code, use the provided functions to modify the current file. Do not include the modified code in your final response unless explicitly asked. If the user is asking you to revert or undo a change, tell them that you are not able to remember past file contents. BRIEFLY EXPLAIN YOUR CHANGES IN ONE SHORT MESSAGE."
        : "If the user is asking you to add, remove, or replace code, explain that you are not able to do that. However, you can provide a code snippet so they can copy and paste it." +
            "\n\nUser: ";
    case "idle":
      return "The user may be stuck on a line of code. Send them a SHORT message to see if they need help. Be sure to include the line of code in your message as a Markdown code block, BUT NOT IF THE LINE IS EMPTY. BRIEFLY EXPLAIN YOUR REASONING TO SEND A MESSAGE.";
    case "completed":
      return `The user has just completed a block of code. You may now respond CONCISELY in one of the following ways:
1. If the completed block is too small or insignificant to comment on, or you have nothing significant to add, respond "NO_RESPONSE".
2. If you spot an issue, notify the user in a SHORT message.
3. If you spot a potential optimization or refactoring operation, suggest it to the user in a SHORT message.
4. If you find documentation opportunities, add comments in editor that fit the code.
If you send a response, BRIEFLY EXPLAIN YOUR REASONING TO SEND A MESSAGE.`;
    case "commented":
      return `The user has just entered a new line after a comment. You may now respond CONCISELY in one of the following ways:
1. If you have nothing significant to add regarding the comment, respond "NO_RESPONSE".
2. If the comment documents code, but there is no code written after the comment, add code that fits the comment.
3. If the comment is posing a question, offer assistance in a SHORT message.
If you send a response, BRIEFLY EXPLAIN YOUR REASONING TO SEND A MESSAGE.`;
    case "multiLineChange":
      return `The user has just made a multiline change. Analyze the change and respond CONCISELY in one of the following ways:
1. If the change is too small or insignificant to comment on, or you have nothing significant to add, respond "NO_RESPONSE".
2. If the change requires documentation, add comments in editor that fit the code.
3. If you spot an issue with the change, notify the user in a SHORT message.
If you send a response, BRIEFLY EXPLAIN YOUR REASONING TO SEND A MESSAGE.`;
    case "selected":
      return `The user has just selected a range of messages. You may now respond CONCISELY in one of the following ways:
1. If the selection is too small or insignificant to comment on, or you have nothing significant to add, respond "NO_RESPONSE".
2. If the selection is a code snippet, spot any errors or ask user if they need help in a SHORT message.
If you send a response, BRIEFLY EXPLAIN YOUR REASONING TO SEND A MESSAGE.`;
    case "breakout":
      return `Your chat with the user is now being split into a separate interface. This interface should only contain messages relevant to the specific task you just completed, including the user message that triggered the task. Please select the range of relevant messages using the selectMessages function. DO NOT EXPLAIN YOUR SELECTION TO THE USER.\n\n`;
    default:
      return "";
}
\end{lstlisting}

\section{Task Descriptions}
\label{appendix:task}
\subsection{Task 1: Scheduling API}

Implement a scheduling system class that maintains a list of events, and provides a method to create new events. You need to check whether there are location or participant conflicts between a new event and created events.

\subsubsection{Subtasks}

\begin{enumerate}
    \item Maintain a list of events, including all related information (name, time, participants, location)
    \item Implement method to add a new event using provided parameters 
    \item Check for location and participant conflicts when adding a new event
    \item Display events in a list, sorted by time
\end{enumerate}

\subsection{Task 2: Word Guessing Game}

Implement a word guessing game (\textit{i.e.} Wordle) using the provided Dictionary API endpoint. The game manager class is initialized with a five-letter word (verified by API). It requires a method to return unguessed letters, and a method for guessing the word, which returns feedback (e.g. \verb|‘???X!’|).

\subsubsection{Subtasks}

\begin{enumerate}
    \item Implement an initialize method that takes an arbitrary string, verify it’s five letters
    \item Store the game state, and implement a method to return the set of unguessed letters
    \item Implement a guess method, which takes a five-character string and returns a feedback string
    \item Use the dictionary API (endpoint provided) to verify if the input word is an actual word, and return an error if not
\end{enumerate}

\subsection{Task 3: Budget Tracker}

Implement a budget tracker class that keeps track of income and spending. The class is initialized with a starting amount. It contains methods to add income and expenses with category and amount, to calculate existing balance, to set spending limits on expense categories, and to create a spending report.

\subsubsection{Subtasks}

\begin{enumerate}
    \item Implement methods that allow users to add sources of income and track expenses, including descriptions and amounts.
    \item Implement a method that calculates the current balance based on the added income and expenses.
    \item Implement a method that enables users to set budget limits for different expense categories, and gives warnings when limits are exceeded.
    \item Implement a method that generates spending reports showing the breakdown of expenses by category.
    \begin{itemize}
        \item The report should display categories with limits first, sorted by the distance from the category limit (ascending). 
        \item Then, the report should display categories without a limit, sorted by the total expense amount (descending).
    \end{itemize}
\end{enumerate}

\end{document}